\documentclass[aps,prx,reprint,nofootinbib,superscriptaddress,longbibliography,floatfix]{revtex4-2}

\usepackage{amsmath,amssymb}
\usepackage{bm}
\usepackage{graphicx}
\usepackage{booktabs}
\usepackage{xcolor}
\usepackage{placeins}
\usepackage[colorlinks=true,allcolors=blue]{hyperref}
\usepackage{comment}
\newcommand{\ev}[1]{\langle #1\rangle}

\newcommand{\PXP}{\textup{PXP}}

\newcommand{\qv}{\bm q}
\newcommand{\Ov}{\bm O}

\begin{document}

\title{Reconstructing hidden observables via quantum snapshot thermometry
}

\author{Jie Ren}
\affiliation{School of Physics and Astronomy, University of Leeds, Leeds LS2 9JT, United Kingdom}

\author{Andrew Hallam}
\affiliation{School of Physics and Astronomy, University of Leeds, Leeds LS2 9JT, United Kingdom}
\affiliation{EPCC, University of Edinburgh, Edinburgh, EH8 9BT, United Kingdom}

\author{Luheng Zhao}
\affiliation{Duke Quantum Center, Duke University, Durham, NC 27701, USA}
\affiliation{Department of Electrical and Computer Engineering, Duke University, Durham, NC 27708, USA}

\author{Huanqian Loh}
\affiliation{Duke Quantum Center, Duke University, Durham, NC 27701, USA}
\affiliation{Department of Electrical and Computer Engineering, Duke University, Durham, NC 27708, USA}
\affiliation{Department of Physics, Duke University, Durham, NC 27708, USA}

\author{Zlatko Papi\'c}
\affiliation{School of Physics and Astronomy, University of Leeds, Leeds LS2 9JT, United Kingdom}

\begin{abstract}
Projective snapshots are a standard readout tool for quantum simulators, but they directly probe only observables that are diagonal in the measurement basis. 
We show how local equilibration and an independently calibrated equilibrium equation of state can reveal otherwise hidden conserved-density profiles.  We demonstrate numerically that such ``quantum snapshot thermometry'' can recover energy fronts and a diffusion constant, distinguish spin transport from energy spreading, and reveal conserved-charge profiles invisible in the initial measurements.
We also identify symmetry-imposed ambiguities in the readout and show how prior knowledge of the spatial structure can resolve them.
Fixed-basis snapshot thermometry thus provides quantitative hydrodynamic information beyond transport exponents, bypassing additional measurement bases and quantum state tomography.
\end{abstract}

\maketitle

\begin{figure}[t]
  \centering
  \includegraphics[width=0.99\columnwidth]{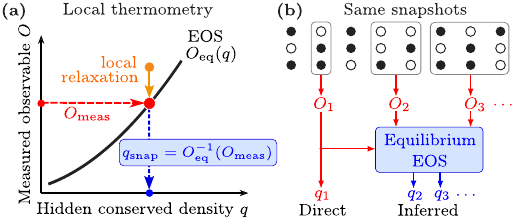}
  \caption{\textbf{Quantum snapshot thermometry principle.}
\textbf{(a)}~The equilibrium EOS, $O_{\rm eq}(q)$,
connects a measured diagonal observable $O$ to a hidden
conserved density $q$. Once local equilibrium is reached at fixed $q$ (orange arrow), the dashed red and blue
arrows illustrate reconstruction via
$q_{\rm snap}=O_{\rm eq}^{-1}(O_{\rm meas})$.
\textbf{(b)}~Multiple hidden conserved densities can be reconstructed from the same snapshots. Local diagonal observables $O_1,O_2,O_3,\ldots$ are measured from the same set of snapshots. In this example, $O_1$ directly determines a conserved
density $q_1$ (red). Based on this input and other measured observables, an equilibrium EOS allows us to infer the remaining  conserved densities $q_2,q_3,\ldots$ (blue). This reconstruction requires a sufficient set of observables with independent responses to the conserved densities.
}
  \label{fig:principle}
\end{figure}

\section{Introduction}

In analog quantum simulators, key information about the many-body wave function is extracted from site-resolved projective measurements~\cite{Bakr2009,Bernien2017,Wei2022,Joshi2022,Rosenberg2024}.
These wave-function snapshots give direct access to observables that are diagonal in the measurement basis.
However, many quantities of interest, such as local energy, contain off-diagonal terms and thus cannot be directly determined from the snapshots.
Such observables can be accessed through additional measurement settings, including basis rotations and randomized measurement protocols~\cite{Huang2020,Elben2023}, which can be challenging on platforms with limited coherent control~\cite{Liu2024HamiltonianShadow}.
Alternatively, off-diagonal observables can be inferred from fixed-basis measurements following known Hamiltonian dynamics~\cite{Hu2022,Liu2024HamiltonianShadow,Peruzzo2026},  although reconstructing them can require computationally expensive processing in an exponentially large Hilbert space. Hamiltonian shadow protocols alleviate this cost by using independently evolving patches~\cite{Liu2024HamiltonianShadow}, while other protocols employ ancillary degrees of freedom~\cite{Tran2023,McGinley2023}. These approaches use dedicated quantum dynamics to encode information into measurement outcomes, motivating a complementary strategy that exploits local equilibration during the dynamics of interest.

Recent work has proposed that principal component analysis (PCA) of wave-function snapshots can diagnose transport and extract the associated dynamical exponents~\cite{Bhakuni2024,Yadav2026}.
The observed correspondence between PCA entropy and energy transport raises the question of how a conserved-density profile becomes encoded in snapshot statistics when the corresponding operator is not diagonal in the measurement basis.
Understanding this mechanism could enable quantitative reconstruction of the evolving profile, including its amplitude and shape, beyond transport exponents. A possible route is suggested by equilibrium thermometry, where equations of state and fluctuation relations connect measurable observables to thermodynamic quantities~\cite{Nascimbene2010EOS,Ku2012Thermodynamics,Hartke2020,Pasqualetti2024}.
Atom-resolved fluctuation measurements have also enabled local thermometry without requiring global thermal equilibrium~\cite{Dixmerias2026}.

Here we show that local equilibration enables quantitative reconstruction of hidden conserved-density profiles from fixed-basis snapshots, without additional measurement operations or global dynamical reconstruction. Once local equilibration is established, an independently calibrated equilibrium equation of state (EOS) relates measurable diagonal observables to the local conserved densities.
Where this relation is locally invertible, it converts snapshot averages into the density profile [Fig.~\ref{fig:principle}(a)].
Note that \emph{global} invertibility is \emph{not} essential: independent information about the preparation and spatial structure of the evolving profile can select the appropriate EOS branch.
We demonstrate that this readout recovers the shape and amplitude of energy fronts and enables extraction of the diffusion constant.
With multiple conserved quantities, different observables extracted from the same snapshots distinguish their transport responses, e.g., spin transport from energy spreading [Fig.~\ref{fig:principle}(b)].
The framework also applies beyond energy-conserving dynamics, enabling reconstruction of conserved-charge profiles in Floquet circuits that are not directly accessible in the measurement basis.

The remainder of this paper is organized as follows.
Section~\ref{sec:local} introduces the local-equilibrium readout, its requirements and sources of error.
Sections~\ref{sec:energy_mfim}-\ref{sec:floquet} apply this approach to numerically demonstrate energy reconstruction in the mixed-field Ising model, joint spin and energy readout in the XXZ chain, and hidden conserved-charge reconstruction in a Floquet circuit, respectively.
Section~\ref{sec:pxp} examines symmetry-imposed ambiguities and branch-resolved reconstruction on the example of blockaded Rydberg atom arrays. Section~\ref{sec:pca} explains the relation of PCA entropy~\cite{Bhakuni2024,Yadav2026} with local spin variance and identifies conditions under which it inherits transport scaling.
Our conclusions are presented in Section~\ref{sec:conc}, while the appendices provide supporting analytical and numerical details.

\section{Local equilibration and readout}\label{sec:local}

Let $\qv(x,t)\equiv \{ q_a(x,t)\}$ denote a set of local conserved
densities, while $\Ov(x,t)\equiv \{ O_a(x,t)\}$ are the diagonal observables accessible
from snapshots.
Local equilibration relates these observables to the
conserved densities through an equilibrium equation of
state (EOS)~\cite{Rigol2008}:
\begin{equation}\label{eq:residualmain}
  O_a(x,t)=O_{a,\mathrm{eq}}[\qv(x,t)]+r_a(x,t),
\end{equation}
where $r_a$ denotes departures from the chosen equilibrium
relation, including the influence of possible degrees of freedom beyond $\qv$. For the example in Fig.~\ref{fig:principle}(a) with a single conserved density,
$r_a$ is the vertical displacement from the equilibrium
curve at a given density $q$, and the orange arrow depicts the process of  local relaxation toward this curve.
Once $r_a$ becomes small, the measured average $O_{\rm meas}$
gives the reconstructed density
\begin{equation}\label{eq:snapshotreadout}
  q_{\rm snap}=O_{\rm eq}^{-1}(O_{\rm meas}),
\end{equation}
on a selected invertible EOS branch. This forms the basis of our thermometry protocol, illustrated by dashed arrows in Fig.~\ref{fig:principle}(a).

The described protocol requires knowledge of $\Ov_{\rm eq}(\qv)$, which should be obtained via an independent calibration. We do this by evaluating the diagonal
observables and conserved densities in uniform equilibrium
states spanning the relevant density range. In this work, we will focus on one-dimensional (1D) systems and finite-temperature regimes, where local equilibrium averages can be evaluated efficiently using thermal matrix product state (MPS) methods, with the system size needed for calibration set by convergence of the local observables. 
However, the thermometry protocol itself is not dimension-specific. In higher dimensions, the EOS could be obtained using other numerical methods or experimental measurements, provided that they associate observable values with independently determined conserved densities. At fixed system parameters, the same EOS can be reused for
different preparations whenever the evolving observables
follow the calibrated relation.

For small deviations from a uniform reference density
$\qv_0$, the linear response is governed by the EOS Jacobian 
\begin{equation}\label{eq:decoder}
  \mathcal{J}_{a\alpha}(\qv)
  =
  \frac{\partial O_{a,\mathrm{eq}}(\qv)}
       {\partial q_\alpha},
\end{equation}
evaluated at $\qv=\qv_0$.
For a single conserved quantity, a nonzero EOS slope
allows readout from one diagonal observable.
For energy readout, this derivative also characterizes
the observable's leading coupling to the slow energy mode,
connecting the EOS response to hydrodynamic
relaxation~\cite{Capizzi2025,Wang2026ETH}.
With several conserved quantities, each observable
generally mixes their contributions.
Resolving all densities $q_a$ at linear order requires
$\mathcal{J}$ to have full column rank, so that the observables
respond independently to changes in the densities.
For equal numbers of observables and densities, this
condition reduces to $\det \mathcal{J}\neq0$.
Furthermore, when linearization about $\qv_0$ is insufficient,
reconstruction should use the full EOS.

Figure~\ref{fig:principle}(b) illustrates the general protocol with multiple hidden conserved densities $q_a$ reconstructed from diagonal observables $O_a$, measured in the same ensemble of snapshots.  A typical example is the joint readout of spin and energy: while magnetization $m$ is diagonal in the standard Pauli basis, energy $e$ contains off-diagonal terms that ordinarily require additional measurement settings to access.
The same snapshot ensemble provides the local
magnetization $m_i=\langle\hat\sigma_i^z\rangle$ and nearest-neighbor correlation $C_i=\langle\hat\sigma_i^z\hat\sigma_{i+1}^z\rangle$. In the notation above, these correspond to $\qv=(m,e)$ and $\Ov=(m,C)$.
Magnetization gives the spin density $m(x,t)$ directly. The independently calibrated relation $C=C_{\rm eq}(e,m)$ then determines the energy $e(x,t)$ from the measured magnetization, provided that $\partial_e C_{\rm eq}\neq0$.

A few comments are in order. First, local invertibility does not imply that the EOS is globally one-to-one. Symmetries can make different densities produce the same measured observables, or force the local response to vanish for particular states. Reconstruction can then require additional observables
or independent input to select an EOS branch, as discussed in 
Sec.~\ref{sec:pxp}. Second, successful readout requires the measured observables to follow the calibrated EOS at the instantaneous local densities. Local equilibration can establish this relation before
transport erases the spatial profile, providing a time
window for its reconstruction~\cite{Lux2014,Kaufman2016Thermalization,Kim2018DetailedBalance,Le2023Hydrodynamization,Shaw2025Ergodicity}.
The initial state need not be thermal, and the evolving
local state need not be fully described by the Gibbs ensemble.
Also, additional conserved quantities can produce
deviations from a Gibbs EOS, whose effect on the readout
must be assessed explicitly in the case of integrable models, as done in Sec.~\ref{sec:xxz}.

To quantify reconstruction errors, we write the measured
snapshot averages at the true local density $\qv$ as
$\widehat{\Ov}=\Ov_{\rm eq}(\qv)+\boldsymbol{r}
+\boldsymbol{\eta}$, where $\boldsymbol{r}$ describes
departures from the equilibrium EOS and
$\boldsymbol{\eta}$ is the fluctuation due to a finite number of samples.
The calibrated EOS may also differ from the exact
equilibrium relation:
$\Ov_{\rm cal}=\Ov_{\rm eq}+\delta\Ov_{\rm cal}$.
The reconstructed density $\qv_{\rm snap}$ is obtained
by matching this calibrated relation to the measured
averages, $\Ov_{\rm cal}(\qv_{\rm snap})=\widehat{\Ov}$.
For a square, invertible Jacobian, expanding this equation
about the true density $\qv$ gives, to leading order
in the errors,
\begin{equation}\label{eq:readouterror}
  \qv_{\rm snap}-\qv
  \simeq
  \mathcal{J}(\qv)^{-1}
  \left(
    \boldsymbol{r}
    +\boldsymbol{\eta}
    -\delta\Ov_{\rm cal}(\qv)
  \right).
\end{equation}
Departures from the EOS and calibration errors therefore
produce a systematic shift in the reconstructed density,
while sampling fluctuations produce statistical
uncertainty.
The inverse Jacobian determines how these observable
errors translate into density errors.
If the observables respond only weakly to density changes, even small observable errors can produce large
reconstruction errors. Thus, as in ordinary thermometry, 
local invertibility alone does not ensure accurate readout.

For a single conserved density reconstructed from one
observable, averaging over $N_{\rm shot}$ independent
snapshots reduces the variance of the measured average
by a factor of $N_{\rm shot}$.
The local EOS slope converts this observable uncertainty
into density uncertainty, giving
\begin{equation}\label{eq:readoutvariance}
  \operatorname{Var}(q_{\rm snap})
  \simeq
  \frac{\operatorname{Var}(O^{(1)})}
       {N_{\rm shot}[O'_{\rm eq}(q)]^2},
\end{equation}
where $O^{(1)}$ is the observable's value in one snapshot
and the derivative is evaluated at the true density $q$.
Reconstruction is therefore more precise when
single-snapshot fluctuations are smaller or the EOS
responds more strongly to changes in density~\cite{DePasquale2016}.
At fixed nonzero EOS slope and single-snapshot variance,
resolving a density contrast $\Delta q$ with the same
relative statistical uncertainty requires
$N_{\rm shot}\propto(\Delta q)^{-2}$. 
Increasing the snapshot count reduces this statistical
uncertainty but does not remove the systematic error
in Eq.~\eqref{eq:readouterror}, which must also remain
small compared with $\Delta q$.
Finally, at a point where the EOS slope vanishes,
Eq.~\eqref{eq:readoutvariance} no longer applies.
If the EOS can still be inverted on a selected branch,
one can instead obtain a density uncertainty interval
by applying that branch's full inverse EOS to the
uncertainty interval of the measured observable.

\begin{figure}[!t]
  \centering
  \includegraphics[width=\columnwidth]{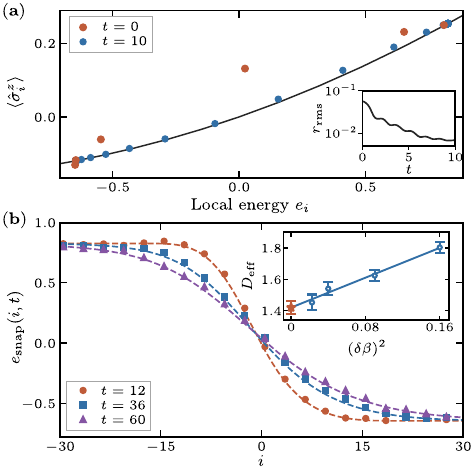}
  \caption{
  \textbf{Energy diffusion from MFIM snapshots.}
  \textbf{(a)}~Local magnetization versus energy density at the
  junction at $t=0$ (orange) and in the bulk at $t=10$ (blue),
  compared with the independently calibrated EOS (black).
  Inset: root-mean-square deviation $r_{\rm rms}$ from the EOS
  over the six sites nearest to the junction.
  \textbf{(b)}~Reconstructed energy profiles at $t=12,36,60$
  (symbols, with $N_{\rm shot}=8192$), compared with
  directly computed profiles (dashed).
  Profiles are smoothed over seven sites, where  $i=0$ denotes the center of the initial wall.
  Inset: effective diffusion constants from the growth of
  reconstructed front widths over $12\le t\le60$.
  The line fit is $D_{\rm eff}=D_0+c(\delta\beta)^2$,
  with the star marking $D_0=1.420\pm0.039$.
  Error bars show one sampling standard deviation estimated
  from $2048$ bootstrap resamples.
  The dynamics are simulated using DMT with bond dimension
  $\chi=256$ on a chain of $L=160$ sites and $\delta\beta=0.2$ in the main plots in (a)-(b).}
  \label{fig:mfim}
\end{figure}

\section{Energy diffusion}\label{sec:energy_mfim}

As our first example, we will demonstrate the reconstruction of energy spreading in the 1D mixed-field Ising model (MFIM):
\begin{equation}\label{eq:MFIM}
  \hat H_{\rm MFIM}
  =J\sum_i\hat\sigma_i^z\hat\sigma_{i+1}^z
  +g_x\sum_i\hat\sigma_i^x
  +g_z\sum_i\hat\sigma_i^z,
\end{equation}
where $\hat\sigma_i^{x,z}$ are the standard Pauli matrices. We fix the Ising coupling $J=1$ and set the field strengths to
$(g_x,g_z)=(1.4,0.9045)$, which places the model far from its integrable limit~\cite{YiThomas2024}. For the Hamiltonian~(\ref{eq:MFIM}), energy is the only conserved density and spreads
diffusively at late times~\cite{Kim2013,White2018,YiThomas2024}.
The local energy $\hat h_i$ in the bulk is
\begin{equation}
  \hat h_i
  =J\hat\sigma_i^z\hat\sigma_{i+1}^z
  +\frac{g_x}{2}
   (\hat\sigma_i^x+\hat\sigma_{i+1}^x)
  +\frac{g_z}{2}
   (\hat\sigma_i^z+\hat\sigma_{i+1}^z),
  \label{eq:mfimeosbond}
\end{equation}
where we have divided each local
field equally between adjacent bonds.
The energy profile to be reconstructed is
$e_i(t)=\ev{\hat h_i}$, which is clearly not diagonal in the computational basis.

Initially, we prepare a thermal domain wall between two Gibbs ensembles,
\begin{equation}
\hat\rho_0=\hat\rho_L\otimes\hat\rho_R,\qquad
  \hat\rho_\alpha\propto e^{-\beta_\alpha\hat H_\alpha},
  \quad \alpha=L,R,
  \label{eq:gibbswall}
\end{equation}
where $\hat H_L$ and $\hat H_R$ act on the disconnected
halves of the chain, see Appendix~\ref{app:formalism} for details.
The inverse temperatures are set according to $\beta_{L,R}=\mp\delta\beta$, so that the wall
probes energy transport about infinite temperature.
Although the reservoirs are initially thermal, the
junction is locally out of equilibrium with respect to
the full Hamiltonian. 
To simulate transport efficiently, tensor-network methods such as density matrix truncation (DMT)~\cite{White2018,YiThomas2024} and dissipation-assisted operator evolution (DAOE)~\cite{Rakovszky2022DAOE,vonKeyserlingk2022Backflow,Srivatsa2026DAOE} retain local information while suppressing the growth of operator entanglement. Here we evolve the state in Eq. (\ref{eq:gibbswall}) under $\hat H_{\rm MFIM}$ using DMT, representing the density matrix as a matrix product operator. During bond-dimension compression, DMT preserves the expectation values of short-range operators, including local magnetization and energy density. 

The transverse-field contribution to the energy cannot
be read directly from $\sigma^z$-basis snapshots.
Once local equilibration is established, however, the
energy density can be inferred from the magnetization
$m_i=\ev{\hat\sigma_i^z}$ through $m_{\rm eq}(e)$, as explained in Sec.~\ref{sec:local}.
We calibrate this EOS independently in uniform Gibbs 
states of an $L=24$ chain. The equilibrium magnetization 
and bond energy are evaluated as functions of inverse temperature, 
we then eliminate the temperature by interpolating the resulting $(e,m)$ pairs. The calibrated curve is monotonic over the energy range
explored by the wall (see Appendix~\ref{app:formalism}), allowing the magnetization to
determine the energy uniquely.
Figure~\ref{fig:mfim}(a) shows that the evolving local
pairs $(e_i,m_i)$ approach this curve after a short
transient, including near the junction which is initially out of equilibrium.

Each $\sigma^z$-basis snapshot yields outcomes
$s_i^{(r)}=\pm1$, where $r$ labels the snapshot.
For $N_{\rm shot}$ independent snapshots at each time step,
the measured magnetization and reconstructed energy are
\begin{equation}
  m_{\rm est,i}(t)
  =\frac{1}{N_{\rm shot}}
   \sum_{r=1}^{N_{\rm shot}}s_i^{(r)}(t),
  \quad
  e_{\rm snap,i}(t)
  =m_{\rm eq}^{-1}[m_{\rm est,i}(t)].
  \label{eq:shotfeatures}
\end{equation}
With a realistic value of $N_{\rm shot}=8192$ snapshots, Fig.~\ref{fig:mfim}(b) shows that the reconstructed profiles reproduce well the directly computed energy profiles.

Previous work used the growth of PCA entropy to extract transport exponents from snapshots~\cite{Bhakuni2024}. Here, resolving the spatial profile also gives access to the shape of the spreading front and the diffusion constant, neither of which is fixed by the exponent. The diffusion constant can be extracted from the growth of the reconstructed energy wall’s squared width $V$,
\begin{equation}
  V(t)\simeq \mathrm{const}+2D_{\rm eff}\, t.
  \label{eq:mfimwidthgrowth}
\end{equation}
To keep sampling costs manageable, we use moderate temperature steps, which also probe temperature-dependent transport~\cite{ZanociSwingle2021}. Reflection symmetry makes the broadening even in $\delta\beta$, allowing us to extrapolate to linear response using
\begin{equation}
  D_{\rm eff}(\delta\beta)
  =D_0+c(\delta\beta)^2+O(\delta\beta^4).
  \label{eq:mfimamplitude}
\end{equation}
We obtain $D_0=1.420\pm0.039$ [inset of Fig.~\ref{fig:mfim}(b)], in agreement with the infinite-temperature benchmark from equilibrium current correlations~\cite{YiThomas2024}.

\begin{figure*}[t]
  \centering
  \includegraphics[width=\textwidth]{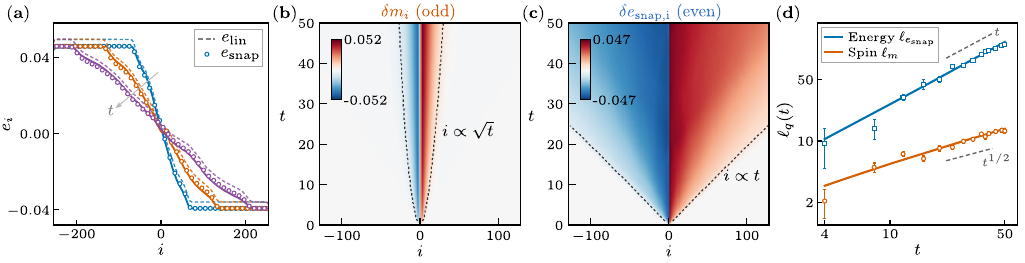}
  \caption{
  \textbf{Spin and energy transport in the XXZ model.}
  \textbf{(a)}~Directly evaluated local energy (solid line), linear
  readout $e_{\rm lin}$ from Eq.~(\ref{eq:xxztwovariableeos}) (dashed line), and  numerical joint-EOS readout based on Eq.~(\ref{eq:xxzmainreadout})   (circles), at $t=12,24,38$ (blue, orange, and purple, respectively). In this panel, the data are obtained by computing expectation values rather
  than sampling, in order to isolate the accuracy of the recovery map from
  sampling noise.
  \textbf{(b)}~Magnetization changes and
  \textbf{(c)}~reconstructed energy changes relative to their
  initial profiles, with the junction located at $i=0$.
  Dashed lines indicate the spreading fronts, consistent with diffusive $\propto \sqrt{t}$ and ballistic $\propto t$ dependence.
  \textbf{(d)}~Normalized integrated changes of spin and
  reconstructed energy, Eq.~(\ref{eq:xxzmainlength}).
  Lines use expectation values of $m,C$; circles and
  squares show spin and reconstructed-energy estimates
  from $32768$ snapshots per time step.
  The energy readout applies the joint EOS to the
  magnetization and correlation averaged over all snapshots.
  Error bars show sampling uncertainties.
  Dashed guides show $t^{1/2}$ and $t$.
  The initial steps are $\delta\beta=0.01$ and
  $\delta\mu=0.1$.
  Dynamics use DMT with bond dimension $\chi=128$ and
  time step $\tau=0.10$ on a chain of $L=512$ sites.
  The ballistic front reaches the boundary near $t\approx 46$.
  }
  \label{fig:xxz}
\end{figure*}

\section{Reading out multiple conserved densities}\label{sec:xxz}

In the MFIM, magnetization served as an energy thermometer. We now show how to read out energy alongside a conserved magnetization, using two observables extracted from the same snapshots. We consider the spin-$1/2$ XXZ chain,
\begin{equation}
  \hat H_\mathrm{XXZ}=\sum_i\bigl(
  \hat\sigma_i^x\hat\sigma_{i+1}^x+
  \hat\sigma_i^y\hat\sigma_{i+1}^y+
  \Delta\hat\sigma_i^z\hat\sigma_{i+1}^z\bigr),
  \label{eq:xxzhamiltonian}
\end{equation}
with anisotropy $\Delta$. The XXZ model conserves total magnetization, $\hat M=\tfrac12\sum_i\hat\sigma_i^z$. While the precise value of anisotropy is unimportant, below we will focus on $\Delta=1.5$, where we expect a clear difference between spin and energy transport: at half filling and near infinite temperature, spin transport is diffusive while energy transport is ballistic~\cite{Zotos1997,KlumperSakai2002,Ljubotina2017}.

We join two Gibbs reservoirs as in Eq.~\eqref{eq:gibbswall}, now with 
\begin{equation}\label{eq:rho0_XXZ}
 \hat\rho_\alpha\propto e^{-\beta_\alpha\hat H_\alpha-\mu_\alpha\hat M_\alpha}, \quad \alpha=L,R,   
\end{equation}
with $\beta_{L,R}=\mp\delta\beta$ and $\mu_{L,R}=\mp\delta\mu$. The same $\sigma^z$-basis snapshots give the magnetization $m_i=\langle\hat\sigma_i^z\rangle$ and nearest-neighbor correlation $C_i=\langle\hat\sigma_i^z\hat\sigma_{i+1}^z\rangle$. Within the uniform Gibbs family, these observables satisfy $C=C_{\rm eq}(e,m)$. 

At half filling and near infinite temperature, the inverse relation gives the energy per bond as 
\begin{equation}
  e_{\rm eq}(m,C)
  =\frac{2+\Delta^2}{\Delta}C
   -\frac{2}{\Delta}m^2
   +O(C^2,Cm^2,m^4),
  \label{eq:xxztwovariableeos}
\end{equation}
which is derived in  Appendix~\ref{app:xxz}. One may be tempted to use the linear term for energy readout, $e_{\rm lin}=((2+\Delta^2)/\Delta)C$. At finite magnetization, however, the $m^2$ correction matters; e.g., at $\beta=0$ the Gibbs state factorizes, giving $C=m^2$ and $e=\Delta m^2$, so the linear readout overestimates the energy by $2m^2/\Delta$. 
Figure~\ref{fig:xxz}(a) compares the reconstructed against the directly computed energy profiles at the finite steps $\delta\beta=0.01$ and $\delta\mu=0.1$.
For reconstruction, we use the full numerical EOS, calibrated independently in uniform $L=128$ Gibbs-MPO states, and invert it at the measured magnetization:
\begin{equation}
  e_{\rm snap}(x,t)
  =e_{\rm eq}\!\bigl(m(x,t),C(x,t)\bigr).
  \label{eq:xxzmainreadout}
\end{equation}
 We see that the joint EOS removes the offset of the linear readout and closely reproduces the evolving energy front.

Figure~\ref{fig:xxz}(b)-(c) shows that
the magnetization spreads approximately diffusively near the junction, while the reconstructed energy spreads ballistically.  To compare their  spreading more quantitatively,
we define an effective transfer length within symmetric
spatial windows on either side of the junction,
\begin{equation}
  \ell_q(t)=
  \frac{\sum_{i>0}\delta q_i(t)-\sum_{i<0}\delta q_i(t)}
       {q_L-q_R}, \quad q=m \; \text{or} \; e_{\rm snap},
  \label{eq:xxzmainlength}
\end{equation}
where $\delta q_i(t)\equiv q_i(t)-q_i(0)$ and $q_{L,R}$
are the initial reservoir values.
The numerator adds the density gained on the right
to that lost on the left.
Dividing by the initial density contrast expresses
this redistribution as a length in chain units,
allowing spin and energy transport to be compared
on the same footing.
For a profile that broadens with a fixed scaling shape,
$\ell_q(t)$ is proportional to the width of the spreading
region.  Figure~\ref{fig:xxz}(d) shows $\ell_m(t)$ and $\ell_{e_{\rm snap}}(t)$ growing approximately as $t^{1/2}$ and $t$, respectively, before boundary effects set in. The snapshot estimates closely follow the curves obtained from expectation values, resolving the different spreading laws from the same set of measurement data.

Unlike the MFIM example considered in Sec.~\ref{sec:energy_mfim}, the 1D spin-$1/2$ XXZ chain is integrable and the additional conserved quantities can cause departures from the Gibbs EOS~\cite{Ilievski2015,Bertini2016}. For a purely thermal weak wall at $\Delta=1.5$, linearized generalized hydrodynamics~\cite{CastroAlvaredo2016,Bertini2016} bounds the energy readout error by approximately $4.7\%$ of the full reservoir energy contrast, see Appendix~\ref{app:xxz} for the derivation and numerical checks. 
The XXZ chain nevertheless provides a particularly clear demonstration of simultaneous readout of distinct transport laws. To establish that the joint-readout principle does not rely on integrability, we additionally demonstrate its application to the nonintegrable spin-$1$ Heisenberg chain in Appendix~\ref{app:spin1}.

\section{Floquet readout with off-diagonal charge}
\label{sec:floquet}

Local equilibrium relations also enable readout of
conserved charges in periodically driven systems or quantum circuits without energy conservation~\cite{DAlessioRigol2014,Khemani2018,Rakovszky2018}. Below we illustrate this on a spin-$1$ Floquet circuit that conserves
$\hat N=\sum_j\hat n_j$, with the charge operator defined as
$\hat n_j=(\hat S_j^x)^2$.
This operator is not diagonal in the $S^z$
measurement basis, but local relaxation makes it accessible
through the EOS. We demonstrate this using two nonthermal preparations
with different charge profiles but identical initial
snapshot distributions.

\begin{figure}[!t]
  \centering
  \includegraphics[width=\columnwidth]{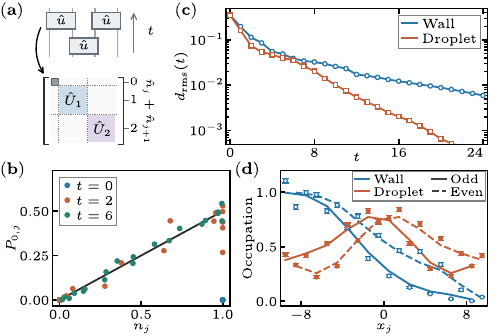}
  \caption{
  \textbf{Readout of hidden charge in a Floquet circuit.}
  \textbf{(a)}~One period of the spin-$1$ brickwall circuit in Eq.~(\ref{eq:floquetbondhamiltonians}),
  applying the same gate $\hat u$ on alternating bonds.
  Each gate conserves the bond occupation $\hat n_i + \hat n_{i+1}$, giving
  blocks of dimensions $1,4,4$ in sectors $0,1,2$.
  \textbf{(b)}~Probability $P_{0,j}$ of a zero
  $S^z$ outcome versus the local charge density
  $n_j=\langle(\hat S_j^x)^2\rangle$ for the wall
  state in Eq.~(\ref{eq:floquetprofiles}).
  Filled circles in blue, orange, and green denote times $t=0,2,6$, respectively, in units of circuit period $2\tau$.
  The black line is the equilibrium EOS, Eq.~(\ref{eq:floquetreadout}).
  \textbf{(c)}~Root-mean-square trace distance $d_{\rm rms}(t)$
  between the one-site reduced states and
  $\hat\rho_{\rm LE}[n_j(t)]$ for the wall
  (blue circles) and droplet (orange squares).
  The quantity $2d_{\rm rms}$ bounds the
  root-mean-square systematic charge-readout error, Eq.~(\ref{eq:floquetreadoutbound}).
  \textbf{(d)}~Charge profiles at $t=6$:
  directly computed densities (solid and dashed
  lines for odd and even sites) and snapshot
  reconstructions (circles), using
  $N_{\rm shot}=2048$ per preparation.
  Both preparations are evolved exactly in an $L=20$ chain
  with $N=10$.
  }
  \label{fig:floquet}
\end{figure}

First, we explain the circuit construction. We will work in the orthonormal local basis, 
$|0_x\rangle =(|+1_z\rangle-|-1_z\rangle)/\sqrt2$, $|0_y\rangle = (|+1_z\rangle+|-1_z\rangle)/\sqrt2$, $|0_z\rangle =|S^z=0\rangle$,
in which the charge operator is diagonal, $\hat n=\operatorname{diag}(0,1,1)$. Let $\hat X,\hat Z$ denote Pauli operators in
the subspace spanned by $\{|0_y\rangle,|0_z\rangle\}$ (we extend the matrices by zero on $|0_x\rangle$).
On a bond $(i,i+1)$, we define the charge transfer 
\begin{equation}
  \hat T
  =\sum_{a=y,z}J_a
  \bigl(|0_x0_a\rangle\langle0_a0_x|
  +\mathrm{h.c.}\bigr).    
\end{equation}
Using $\hat T$, we define the elementary circuit gate $\hat u=\exp(-i\tau\hat H)$, in terms of the bond Hamiltonian:
\begin{equation}
    \hat H=\hat T+K(\hat X_i\hat X_{i+1}+\hat Z_i\hat Z_{i+1})
      +\tfrac{\Omega}{2}(\hat X_i+\hat X_{i+1}).
  \label{eq:floquetbondhamiltonians}
\end{equation}
During each period, we apply $\hat u$ in a standard brickwall pattern,  i.e., first on bonds
$(2,3),(4,5),\ldots$, followed by
bonds $(1,2),(3,4),\ldots$, as shown in 
Fig.~\ref{fig:floquet}(a). The precise values of the parameters are unimportant and below we choose $J_y=0.6$, $J_z=1$, $K=1.1$, $\Omega=0.9$, $\tau=1$. Importantly, the gate $\hat u$ conserves $\hat n_i+\hat n_{i+1}$.

At fixed mean occupation $n$, the 
one-site state with maximum entropy is
\begin{equation}
  \hat\rho_{\rm LE}(n)
  =\operatorname{diag}(1-n,n/2,n/2)\,.
  \label{eq:floquetsuppensemble}
\end{equation}
Since a zero outcome in an $S^z$ snapshot projects
onto $|0_z\rangle$, its equilibrium probability is
\begin{equation}
  P_{0,\rm eq}(n)=n/2.
  \label{eq:floquetreadout}
\end{equation}
The charge constraint and the equal equilibrium
populations of the occupied doublet therefore provide analytic calibration for the readout.

We consider two initial states with the same total charge but different spatial profiles, dubbed the ``wall" and ``droplet",
\begin{equation}
  \begin{aligned}
    |\Psi_{\rm wall}\rangle
    &=
    |0_y\rangle^{\otimes L/2}
    \otimes|0_x\rangle^{\otimes L/2},\\
    |\Psi_{\rm drop}\rangle
    &=
    |0_x\rangle^{\otimes L/4}
    \otimes|0_y\rangle^{\otimes L/2}
    \otimes|0_x\rangle^{\otimes L/4}.
  \end{aligned}
  \label{eq:floquetprofiles}
\end{equation}
Both have $N=L/2$.
The local states $|0_x\rangle$ and $|0_y\rangle$
have $\hat n$ occupations zero and one, respectively, but
both yield $S^z=\pm1$ with equal probability.
Consequently, the two preparations have identical
full initial snapshot distributions despite their
different charge profiles.
In particular, $P_0=0$ everywhere initially, so the
equilibrium readout does not recover the occupied
regions.

As the internal degrees of freedom relax, the measured
zero-outcome probabilities $P_{0,j}(t)$ approach
$n_j(t)/2$ [Fig.~\ref{fig:floquet}(b)].
As a stronger test of local equilibration, we compare
the full one-site reduced density matrix with the
local equilibrium state at the same charge density,
defined in Eq.~\eqref{eq:floquetsuppensemble}, 
\begin{equation}
  d_j(t)=\frac12
  \left\|
  \hat\rho_j(t)-\hat\rho_{\rm LE}[n_j(t)]
  \right\|_1\,,
  \label{eq:floquettracedistance}
\end{equation}
with $\|\ldots \|_1$ being the trace norm.
For both preparations, the root-mean-square distance
$d_{\rm rms}(t)=[L^{-1}\sum_jd_j(t)^2]^{1/2}$
decreases with time, as seen in Fig.~\ref{fig:floquet}(c).
This also provides a quantitative bound on the readout
error: since trace distance bounds the difference
between measurement probabilities,
\begin{equation}
  |2P_{0,j}(t)-n_j(t)|\leq2d_j(t).
  \label{eq:floquetreadoutbound}
\end{equation}
Thus, $2d_{\rm rms}$ bounds the root-mean-square
error due to incomplete local equilibration,
before sampling noise is included.
With finite sampling, we reconstruct the density as
$n_{{\rm snap},j}=2N_{0,j}/N_{\rm shot}$, where
$N_{0,j}$ counts zero outcomes at site $j$.
Figure~\ref{fig:floquet}(d) shows that $2048$ snapshots
per preparation recover both the wall and droplet
profiles at $t=6$.
Local relaxation thus reveals the initially hidden
charge profiles while they retain substantial spatial
structure.

\begin{figure*}[!t]
  \centering
  \includegraphics[width=\textwidth]{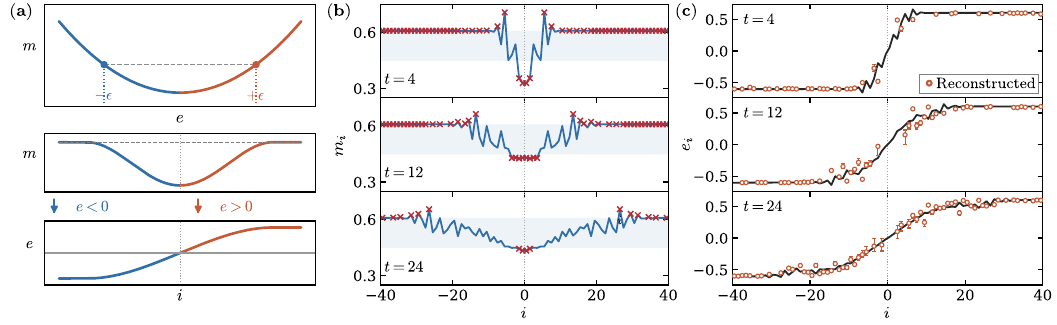}
  \caption{
  \textbf{Branch-resolved energy reconstruction in the PXP model.}
  \textbf{(a)}~Schematic of 
  the magnetization EOS which is an even function of $e$, and branch selection
  from the known wall orientation (see text).
  \textbf{(b)}~Magnetization profiles at $t=4,12,24$.
  Shading marks the calibrated magnetization interval;
  red crosses indicate values outside this interval,
  for which reconstruction is not performed.
  \textbf{(c)}~Directly computed (black) and reconstructed
  (orange) energy profiles. Reconstruction uses magnetization
  averages from $8192$ snapshots per time.
  To prepare a wall, we use the ground states prepared by
  DMRG with $+\hat H$ on the left and $-\hat H$ on
  the right in system size $L=128$. We omit the Hamiltonian terms that straddle the junction during preparation. The subsequent 
  evolution is performed by two-site TDVP with bond dimension $\chi=256$ and time step $\delta t=0.05$.
  The EOS is obtained from Gibbs states of a periodic
  $L=16$ PXP chain over the interval $-12\leq\beta\leq12$, which we interpolate separately on each monotonic branch.}
  \label{fig:pxpcombined}
\end{figure*}

\section{Symmetry constraints and branch-resolved reconstruction}
\label{sec:pxp}

Symmetries can make equilibrium readout globally
ambiguous even after local equilibration.
We illustrate this limitation in the kinetically
constrained \PXP{} model~\cite{FendleySachdev,Lesanovsky2012}, which describes resonantly
driven Rydberg atom chains in the nearest-neighbor
blockade limit~\cite{Browaeys2020,Bernien2017,Bluvstein2021,Zhao2024Observationquantumthermalizationrestricted} (the same model can also be realized in a tilted Bose-Hubbard optical lattice~\cite{Su2023}). In this case, the site-resolved
measurements of atomic state occupations provide the
$Z$-basis snapshots considered below.

The PXP Hamiltonian is
\begin{equation}
  \hat H=\sum_i\hat P_{i-1}\hat X_i\hat P_{i+1},
  \qquad
  \hat P_i=\frac{1+\hat Z_i}{2},
  \label{eq:pxpsmhamiltonian}
\end{equation}
where $\hat X_i,\hat Z_i$ are Pauli operators and
$\hat P_i$ projects onto the atomic ground state. This projector encodes a kinetic constraint: each atom can be excited only if its nearest neighbors are both in the ground state. The constraint makes the model nonintegrable, giving rise to much interesting physics including  quantum many-body
scars~\cite{Turner2018,Ho2019,Lin2019exact,Iadecola2019,Choi2019,Khemani2019,Surace2020,Omiya2022,Kerschbaumer2025} and 
superdiffusive energy transport~\cite{Ljubotina2023,Bhakuni2026,Jiang2026}. Here we test equilibrium readout during the
relaxation of an energy wall, Fig.~\ref{fig:pxpcombined}(a).
The local energy
$e_i=\langle\hat P_{i-1}\hat X_i\hat P_{i+1}\rangle$
is off-diagonal in the measurement basis, whereas
the magnetization $m_i=\langle\hat Z_i\rangle$
is directly accessible.

The PXP model has a ``particle-hole'' symmetry $\hat \Gamma\hat H\hat \Gamma=-\hat H$, generated by $\hat \Gamma=\prod_i\hat Z_i$~\cite{Turner2018PRB,Schecter2018}. The Gibbs states at opposite inverse temperatures
therefore obey
$\hat\rho_{-\beta}=\hat \Gamma\hat\rho_\beta\hat \Gamma$.
Since $\hat \Gamma$ is diagonal in the measurement basis,
their snapshot distributions coincide:
\begin{equation}
  p_{-\beta}(\mathbf s)
  =\langle\mathbf s|\hat\rho_{-\beta}|\mathbf s\rangle
  =\langle\mathbf s|\hat\rho_\beta|\mathbf s\rangle
  =p_\beta(\mathbf s).
  \label{eq:pxpsnapshotambiguity}
\end{equation}
These states have opposite energies, so the energy
sign cannot be determined even with arbitrarily many
snapshots and access to all diagonal correlations.
In particular, the magnetization EOS is even:
\begin{equation}
  m_{\rm eq}(-e)=m_{\rm eq}(e).
  \label{eq:pxpsmeveneos}
\end{equation}
Its two branches therefore give the same measured
magnetization, as illustrated in Fig.~\ref{fig:pxpcombined}(a).

However, independent information about the spatial profile
can select the appropriate branch.
Let $g(e)=m_{\rm eq}(e)$ denote the positive-energy
branch, monotonic over the calibrated interval.
Local equilibrium gives
$m(x,t)\simeq g(|e(x,t)|)$, allowing the measured
magnetization to determine the energy magnitude.
For the wall considered here, the preparation fixes
negative energy on the left and positive energy
on the right [Fig.~\ref{fig:pxpcombined}(a)].
Provided that the evolving profile retains a single
zero crossing at the wall center $x=0$, this
information determines the sign:
\begin{equation}
  e_{\rm rec}(x,t)
  =\operatorname{sgn}(x)\,
   g^{-1}\!\left[m(x,t)\right].
  \label{eq:pxpsmreconstruction}
\end{equation}
The EOS thus supplies the magnitude, while the wall
orientation and assumed zero-crossing structure
supply the EOS branch, see Fig.~\ref{fig:pxpcombined}(a).

With the knowledge of the branch structure of the EOS, we can perform energy reconstruction during the wall's evolution, summarized in
Fig.~\ref{fig:pxpcombined}(b)-(c).
The EOS of the PXP model is calibrated independently in uniform Gibbs
states on a small chain of $L=16$ atoms with periodic boundary conditions. At later times, the reconstructed energy follows the
directly computed profile over much of the calibrated
range, although local deviations remain.
Magnetization values outside that range have no
inverse under the calibrated EOS and are excluded,
as indicated by the red crosses. We note that the EOS, being an even function of $e$, also has a vanishing slope at $e=0$,
making inversion poorly conditioned near the junction.
Small departures from the EOS or finite-sampling
fluctuations can therefore produce comparatively
large energy errors, as discussed in Sec.~\ref{sec:local}. The PXP example thus illustrates two distinct aspects
of the readout: prior spatial information enables
reconstruction from a globally ambiguous EOS,
while local EOS sensitivity controls error amplification.

\section{Snapshot entropy and transport}
\label{sec:pca}

Previous sections showed how to reconstruct conserved-density
profiles from local observable expectation values obtained by snapshots. Using these insights, we are now in a position to explain why the ``PCA entropy''~\cite{Bhakuni2024,Yadav2026}, i.e., the Shannon entropy associated with the principal component analysis (PCA) of snapshot data, can inherit the same scaling as dictated by
transport exponents.
We first establish a statistical relation between
PCA entropy and the total local spin variance.
An equilibrium EOS then connects this variance to
the spreading of the conserved density.

Consider $N_{\rm shot}$ snapshots of a spin-$1/2$
chain with $L$ sites, taken at time $t$ and arranged as rows of a matrix $X$, with entries $X_{si}=\pm1$ representing the outcome at site $i$ in snapshot $s$.  The squared singular values of $X/\sqrt{N_{\rm shot}L}$ are the eigenvalues of
$X^T X/(N_{\rm shot}L)$. In the limit of an infinite number of (independent) snapshots, each matrix element approaches the following quantum correlator:
\begin{equation}
\begin{aligned}
  \left[\frac{X^T X}{N_{\rm shot}L}\right]_{ij}
  &=\frac{1}{N_{\rm shot}L}
    \sum_{s=1}^{N_{\rm shot}}X_{si}X_{sj}\\
  &\xrightarrow{N_{\rm shot}\to\infty}
\frac1L\langle\hat\sigma_i^z\hat\sigma_j^z\rangle_t \equiv \Sigma_{ij}.
\end{aligned}
\label{eq:pcacorrelation}
\end{equation}
The eigenvalues $\lambda_k^2$ of this matrix sum to one and define
the following PCA entropy~\cite{Bhakuni2024,Yadav2026}
\begin{equation}
  S_{\rm PCA}(t)
  =-\sum_{k=1}^L\lambda_k^2\ln\lambda_k^2.
  \label{eq:pcasmdefinition}
\end{equation}

Write $m_i=\langle\hat\sigma_i^z\rangle$ and
$G_{ij}=\langle\hat\sigma_i^z\hat\sigma_j^z\rangle-m_im_j$ (these quantities are generally time-dependent).
The diagonal entries $d_i=G_{ii}=1-m_i^2$ are the
local spin variances, with total
$\mathcal V=\operatorname{Tr}G$.
For $\mathcal V>0$, assume
\begin{equation}
  \frac{\operatorname{Tr}(G^2)}{\mathcal V}\leq K,\qquad
  -\frac{\sum_i d_i\ln d_i}{\mathcal V}\leq C,
  \label{eq:pcasmconditions}
\end{equation}
where $K,C$ are independent of system size and time.
Under these conditions, the main result of this section is
\begin{equation}
  S_{\rm PCA}(t)
  =\frac{\ln L}{L}\mathcal{V}(t)
  +O\!\left(\frac{\mathcal{V}(t)}{L}\right).
  \label{eq:pcamain}
\end{equation}
Thus the leading entropy is proportional to the
total local variance, with a size-dependent
prefactor.

Let us outline the derivation of Eq.~(\ref{eq:pcamain}), which will also yield an explicit error bound.
For $0<\mathcal V<L$, separate $\Sigma$ in Eq.~(\ref{eq:pcacorrelation}) as follows:
\begin{equation}
  \begin{aligned}
    \Sigma&=(1-p)\rho_m+p\rho_G,
    \quad p=\mathcal V/L,
  \end{aligned}
  \label{eq:pcasm-mixture}
\end{equation}
where we have introduced the following matrices with unit trace:
\begin{equation}
  \begin{aligned}
    \rho_m&=\bm m\bm m^T /(L-\mathcal V),
    \quad
    \rho_G=G/\mathcal V.
  \end{aligned}
  \label{eq:rhom_rhoG}
\end{equation}
Additionally, $\rho_m$ has rank one. Thus, Eq.~(\ref{eq:pcasm-mixture}) separates the mean magnetization profile from the connected fluctuations.

Entropy-of-mixing bounds and joint convexity of
relative entropy~\cite{NielsenChuang2010} then give the bound:
\begin{equation}
  A-\frac1L\sum_i h_2(d_i)
  \leq S_{\rm PCA}\leq A,
  \label{eq:pcasmentropybound}
\end{equation}
where we introduced $A=h_2(p)+pS(\rho_G)$,
$h_2(v)=-v\ln v-(1-v)\ln(1-v)$, and $S(\rho)=-\operatorname{Tr}(\rho\ln\rho)$. The assumptions in Eq.~\eqref{eq:pcasmconditions}
imply
$\ln\mathcal V-\ln K_*\leq S(\rho_G)\leq\ln\mathcal V+C$,
where $K_*=\max\{1,K\}$, and
$L^{-1}\sum_i h_2(d_i)\leq(C+1)p$. Substitution into Eq.~\eqref{eq:pcasmentropybound}
combines the fluctuation entropy with the mixing entropy
through $-p\ln p+p\ln\mathcal V=p\ln L$.
Using $0\leq-(1-p)\ln(1-p)\leq p$ then yields
\begin{equation}
  \left|
  S_{\rm PCA}-\frac{\mathcal V}{L}\ln L
  \right|
  \leq
  (C+1+\ln K_*)\frac{\mathcal V}{L}.
  \label{eq:pcasmproportionality}
\end{equation}
This establishes Eq.~\eqref{eq:pcamain}, with a
relative correction of order $1/\ln L$.
The endpoints follow directly: $\mathcal V=0$ gives
zero entropy, while $\mathcal V=L$ gives $\Sigma=G/L$.

We next comment on the conditions in Eq.~(\ref{eq:pcasmconditions}) and what can invalidate them.
The first condition holds for uniformly summable
connected correlations and excludes macroscopic
collective fluctuations. For example, equally likely
all-up and all-down snapshots have maximal total local
variance, $\mathcal V=L$, but zero PCA entropy:
all fluctuations occupy a single collective mode.
In this case,
$\operatorname{Tr}(G^2)/\mathcal V=L$,
violating the first condition.
The second condition controls the contribution of sites
with very small local variance. For independent spins
with $d_i=\varepsilon$, the first ratio equals
$\varepsilon$, whereas the second equals
$\ln(1/\varepsilon)$. Thus, the second bound need not
hold uniformly if the local fluctuations become
increasingly weak with system size or time.
These conditions are sufficient; violating either one
does not by itself imply failure of Eq.~(\ref{eq:pcamain}).

The conditions above establish a statistical relation
between PCA entropy and the total local spin variance, without assuming local equilibrium. To connect this result to transport, we now invoke local equilibrium: an equilibrium EOS relates the local spin variance to the spreading conserved density. Consider a conserved-density front
$q(x,t)=F(x/t^{1/z})$ joining reservoirs with zero
local spin variance, described by some universal function $F$.
The EOS gives
$m(x,t)\simeq m_{\rm eq}[q(x,t)]$ and
\begin{equation}
  \begin{aligned}
    \mathcal V(t)&\simeq t^{1/z}\mathcal I,
    \quad \mathcal I
    {=}{\int} du\,\left[1-m_{\rm eq}(F(u))^2\right],
  \end{aligned}
  \label{eq:pcavariancescaling}
\end{equation}
assuming convergence of the rescaled variance profile under the spatial integral, including its
tails. Combining Eqs.~\eqref{eq:pcasmproportionality}
and~\eqref{eq:pcavariancescaling} gives
\begin{equation}
  S_{\rm PCA}(t)
  \simeq\frac{\ln L}{L}\,\mathcal I\,t^{1/z}.
  \label{eq:pcatransportscaling}
\end{equation}
PCA entropy therefore inherits the transport exponent
to leading order in the regime
$1\ll t^{1/z}\ll L$, provided that corrections to the
profile scaling and the $1/\ln L$ relative correction
to the entropy-variance relation are small.

The transport argument above assumes reservoirs with
zero local spin variance, so that $\mathcal V(t)$
comes entirely from the spreading front.
For reservoirs with nonzero variance, including the
PXP energy wall in Sec.~\ref{sec:pxp}, sites far from
the front already contribute an $O(L)$ background
to $\mathcal V$.
The bound in Eq.~\eqref{eq:pcasmproportionality}
then permits an $O(1)$ remainder in the entropy,
which can exceed the change caused by the spreading
front and requires separate analysis. 

\section{Conclusions}\label{sec:conc}

We have shown how local equilibration and an independently calibrated EOS enable reconstruction of conserved-density profiles from fixed-basis snapshots, even when the corresponding operators contain off-diagonal terms. Our numerical examples recover energy fronts and a diffusion constant, distinguish diffusive spin transport from ballistic energy spreading, and reveal hidden conserved-charge profiles in a Floquet system. We quantify how weak EOS responses and departures from Gibbs equilibrium affect the readout, and show how prior information about the preparation and spatial structure can resolve symmetry-imposed ambiguities. We also relate PCA entropy to the local spin variance, clarifying when it can serve as a transport diagnostic. These results establish snapshot thermometry as a quantitative probe of transport beyond scaling exponents.

Looking ahead, snapshot thermometry could broaden the range of conserved quantities accessible from existing snapshot measurements in experiments on Rydberg atom arrays~\cite{Bernien2017,Datla2026Statistical}, optical lattices~\cite{Wei2022}, trapped ions~\cite{Joshi2022}, and superconducting qubits~\cite{Rosenberg2024}. In such applications, the readout should be tailored to the measurement capabilities of each platform. The EOS response and snapshot fluctuations provide quantitative criteria for selecting observables that maximize reconstruction precision at a fixed sampling budget, offering a route towards systematic, experiment-specific optimization.

On the theoretical side, a natural extension is to develop calibrations based on generalized Gibbs ensembles or equilibrium within dynamically accessible sectors, where conventional Gibbs relations may be insufficient~\cite{Ilievski2015,Zhao2024Observationquantumthermalizationrestricted}. The coupling of nonconserved observables to conserved densities and their products connects this readout mechanism to hierarchical relaxation~\cite{Wang2026Hierarchical} and hydrodynamic descriptions of higher-order correlators~\cite{Wang2026ETH,mishra2026theoryoutoftimeorderedtransport}. These connections motivate extending snapshot thermometry from mean density profiles to fluctuations and spatial correlations of transported quantities, with the prospect of testing dynamical universality through higher-order cumulants~\cite{Rosenberg2024}.

\begin{acknowledgments}
Z.P. would like to thank Kartiek Agarwal for useful discussions. J.R., A.H. and Z.P. acknowledge support by the Leverhulme Trust Research Leadership Award RL-2019-015, and EPSRC Grants EP/Z533634/1, UKRI13780. A.H. also acknowledges support from EPSRC through the Quantum Advantage Turbo Charger (QATCH) grant [UKRI4257].
The numerical calculations were performed using the open-source Julia
packages available via GitHub at \url{https://
github.com/jayren3996}. Computational portions of this work were undertaken on the AIRE HPC system at the University of Leeds. This research was supported in part by grant NSF PHY-2309135 to the Kavli Institute for Theoretical Physics (KITP). L.Z. and H.L. acknowledge support from the Army Research Office (W911NF-26-1-A044) and from the Alfred P. Sloan Foundation.
\end{acknowledgments}

\appendix

\section{Equation of state, local equilibration and energy reconstruction from snapshots}
\label{app:formalism}

Here we provide more details about the thermometry protocol, focusing on the MFIM with the bulk local energy in Eq.~\eqref{eq:mfimeosbond}.
At the physical ends of the open chain, each boundary
field is assigned entirely to its adjacent bond, so that
$\sum_{i=-L/2+1}^{L/2-1}\hat h_i=\hat H_{\rm MFIM}$.

\begin{figure}[t]
  \centering
  \includegraphics[width=\columnwidth]{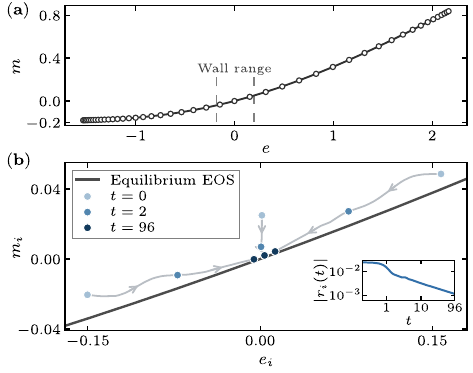}
  \caption{
  \textbf{EOS and local equilibration
  in the MFIM}.
  \textbf{(a)}~Uniform Gibbs-state values for $L=24$
  (circles) and their linear interpolation. Dashed
  guides mark the range of local energies in the
  simulated profiles.
  \textbf{(b)}~Thermal domain wall evolution for $L=160$ and
  $\delta\beta=0.05$.
  Local $(e_i,m_i)$ pairs at
  $i=-1,0,1$ are shown at $t=0,2,96$,
  from light to dark blue, together with the EOS.
  Pale curves trace their evolution, while arrows indicate
  increasing time.
  Inset: the residual $|r_0(t)|$.
  Dynamics use DMT with $\chi=256$ and fourth-order
  time steps~\cite{Suzuki1990} $\tau=0.20$, with $J=1$ and  $(g_x,g_z)=(1.4,0.9045)$.}
  \label{fig:mfims1}
\end{figure}

For the EOS calibration, we prepare uniform Gibbs states
at different inverse temperatures,
$\hat\rho_\beta
  =e^{-\beta\hat H_{\rm MFIM}}/
  {\operatorname{Tr}
   (e^{-\beta\hat H_{\rm MFIM}})}$. 
The states are represented as matrix-product states in
operator space~\cite{Verstraete2004,ZwolakVidal2004}
on an $L=24$ chain.
We evaluate the central bond ($c=0$) energy and
the mean magnetization of its two sites:
\begin{equation}
  \begin{aligned}
    e_{\rm eq}(\beta)
    =\operatorname{Tr}(\hat\rho_\beta\hat h_c),\quad
    m_{\rm eq}(\beta)
    =\frac12\operatorname{Tr}
      \left[\hat\rho_\beta
      (\hat\sigma_c^z+\hat\sigma_{c+1}^z)\right].
  \end{aligned}
  \label{eq:mfimeoscentral}
\end{equation}
Linear interpolation of the resulting $(e,m)$ pairs
gives the smooth calibration curve shown in
Fig.~\ref{fig:mfims1}(a).

The initial energy wall at position $c=0$ on an open chain of even
length $L$ is given by: 
\begin{equation}
  \hat\rho_0\propto
  \exp\!\left[-\sum_{i=-L/2+1}^{L/2-1}\beta_i\hat h_i\right],
  \quad
  \beta_i=
  \begin{cases}
    -\delta\beta,&i<c,\\
    0,&i=c,\\
    +\delta\beta,&i>c.
  \end{cases}
  \label{eq:mfimsmdomainwall}
\end{equation}
This is the same form as Eq.~\eqref{eq:gibbswall}, with $\hat H_L=\sum_{i<c}\hat h_i$ and
$\hat H_R=\sum_{i>c}\hat h_i$. Omitting $\hat h_c$ also removes its half-field contributions, so the fields on the two junction sites
are initially halved. The subsequent evolution is performed with the full Hamiltonian $\hat H_{\rm MFIM}$ in Eq.~(\ref{eq:MFIM}).

We test the calibrated EOS during evolution through
the residual
$r_i(t)=m_i(t)-m_{\rm eq}[e_i(t)]$.
Figure~\ref{fig:mfims1}(b) shows the trajectories of
three sites near the junction for $\delta\beta=0.05$.
The pairs that are initially off-curve move substantially closer
to the EOS by $t=2$, and the residual drops to $10^{-3}$ by $t=96$.

We extract the diffusion constant in
Fig.~\ref{fig:mfim}(b) from the growth of the
reconstructed energy-front width.
To suppress sampling noise,
we restrict the analysis to sites within the distance $|i-\tfrac12|\le50$ from the initial junction.
Before EOS inversion, we fit the measured
magnetizations to a monotonic profile bounded by
the calibrated reservoir values, using weights based
on the estimated sampling variances.
Inverting the EOS gives the fitted energy profile
$\widetilde e_i(t)=m_{\rm eq}^{-1}[\widetilde m_i(t)]$,
which we normalize as
\begin{equation}
  u_i(t)=\frac{\widetilde e_i(t)-(e_L+e_R)/2}{(e_L-e_R)/2}.
  \label{eq:mfims3normalized}
\end{equation}
The squared front width is
\begin{equation}
  V(t)=\sum_{|i-\frac12|\le50}(i-\tfrac12)\,u_i(t)+\mathrm{const.},
  \label{eq:mfimdlit}
\end{equation}
where the constant sets $V=0$ for an ideal sharp wall.
For fixed reservoir plateaus, this expression measures
the second spatial moment of the normalized front
gradient.

\begin{figure}[!h]
  \centering
  \includegraphics[width=\linewidth]{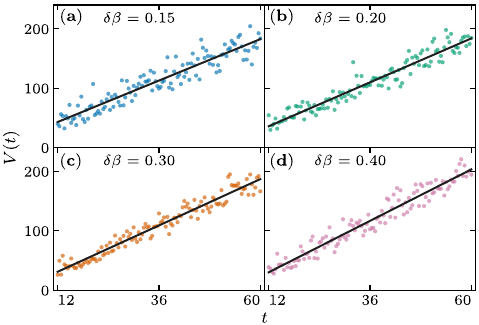}
  \caption{
  \textbf{Moment broadening from MFIM snapshots.}
  Panels (a)--(d) show $V(t)$ at
  $\delta\beta=0.15,0.20,0.30,0.40$, respectively,
  using $8192$ snapshots per time.
  Black lines show $\mathrm{const}+2D_{\rm eff}t$ over
  $12\le t\le60$.
  The chain has $L=160$ sites and bond dimension
  $\chi=256$.}
  \label{fig:mfims2}
\end{figure}

At each wall amplitude, we fit the width growth to
Eq.~\eqref{eq:mfimwidthgrowth} over $12\le t\le60$
to obtain $D_{\rm eff}$.
Figure~\ref{fig:mfims2} shows the broadening for
the four amplitudes, with $8192$ snapshots per profile.
We then obtain $D_0$ by fitting the estimates at
$\delta\beta=0.15,0.20,0.30,0.40$ to
Eq.~\eqref{eq:mfimamplitude}, weighted by their
inverse sampling variances. Sampling uncertainties are estimated from $2048$ bootstrap resamples~\cite{Efron1979Bootstrap}.
We resample whole snapshots independently at each
time and wall amplitude, and repeat the full analysis,
including the extrapolation to $D_0$.
The standard deviations of the resulting diffusion
estimates give the reported uncertainties.
The EOS calibration is held fixed, so these
uncertainties account only for finite sampling.

\section{Gibbs equation of state and error analysis for the XXZ readout}
\label{app:xxz}

At $\beta=0$, the uniform Gibbs state in
Eq.~\eqref{eq:rho0_XXZ} factorizes into single-site
density matrices,
$\hat\rho_i=(\hat I+m\hat\sigma_i^z)/2$, with
$m=-\tanh(\mu/2)$.
Consequently, $C=m^2$ and $e=\Delta m^2$.
Expanding the Gibbs averages using infinite-temperature
Pauli traces gives
\begin{equation}
  \begin{aligned}
    C&=m^2-\Delta\beta-\beta^2
       +O(\beta^3,\beta m^2),\\
    e&=\Delta m^2-(2+\Delta^2)\beta-3\Delta\beta^2
       +O(\beta^3,\beta m^2).
  \end{aligned}
  \label{eq:xxzHighTExpansion}
\end{equation}
Spin-flip symmetry ensures that corrections to the
thermal coefficients begin at order $m^2$.
Eliminating $\beta$ yields
\begin{equation}
  \begin{aligned}
    e_{\rm eq}(m,C)
    &=\frac{2+\Delta^2}{\Delta}C-\frac{2}{\Delta}m^2\\
    &\quad+\frac{2(1-\Delta^2)}{\Delta^3}C^2
    +O(C^3,Cm^2,m^4).
  \end{aligned}
  \label{eq:xxzHighTEOS}
\end{equation}
Keeping the linear term in $C$ and the leading $m^2$
correction gives Eq.~\eqref{eq:xxztwovariableeos}.
At $m=0$, the linear relation is $C=ae+O(e^2)$,
with $a=\Delta/(2+\Delta^2)$.

We assess the error of the linear Gibbs readout
$e_{\rm lin}=C/a$ for a purely thermal weak wall,
$\beta_{L,R}=\mp\delta\beta$ and $\mu_L=\mu_R=0$,
using linearized generalized hydrodynamics
(GHD)~\cite{CastroAlvaredo2016,Bertini2016}.
The calculation describes the Euler-scale profile at
fixed ray $\xi=x/t$, before boundary reflections.
Let $\delta e_\lambda$ and $\delta C_\lambda$ denote
the contributions of quasiparticle mode $\lambda$
to the right reservoir's thermal response.
The mode sum includes all quasiparticle species and
integration over their rapidities.
The mismatch
$r_\lambda=\delta C_\lambda-a\,\delta e_\lambda$
satisfies
\begin{equation}
  \begin{aligned}
    \sum_\lambda\delta e_\lambda
    &=-(2+\Delta^2)\delta\beta,\\
    \sum_\lambda\delta C_\lambda
    &=-\Delta\delta\beta,\qquad
    \sum_\lambda r_\lambda=0.
  \end{aligned}
  \label{eq:xxzThermalSumRules}
\end{equation}
Thus, the Gibbs relation holds after summing over
modes, although individual modes can violate it.

In linearized GHD, a mode at ray $\xi$ comes from the
left reservoir when $v_\lambda>\xi$ and from the
right otherwise.
Opposite velocities have equal thermal response
coefficients.
At $\xi>0$, their contributions therefore cancel
when $|v_\lambda|>\xi$, leaving
\begin{equation}
  \frac{e_{\rm lin}(\xi)-e(\xi)}{\Delta e_{\rm LR}}
  =\frac{1}{a\Delta e_{\rm LR}}
   \sum_{|v_\lambda|<\xi}r_\lambda,
  \label{eq:xxzGHDerror}
\end{equation}
where $\Delta e_{\rm LR}=2\delta\beta(2+\Delta^2)$.
Since the total mismatch vanishes, the magnitude of
any partial sum is at most
$\tfrac12\sum_\lambda|r_\lambda|$.
Using reflection symmetry for negative rays:
\begin{equation}
  \begin{aligned}
    \sup_\xi
    \frac{|e_{\rm lin}(\xi)-e(\xi)|}{\Delta e_{\rm LR}}
    \leq\mathcal{B},
 \quad   \mathcal{B}
    \equiv\frac{1}{2a\Delta e_{\rm LR}}
      \sum_\lambda|r_\lambda|.
  \end{aligned}
  \label{eq:xxzGHDerrorBound}
\end{equation}

We evaluate the mode responses by linearizing the
thermodynamic Bethe-ansatz equations and the
nearest-neighbor correlator formula of
Ref.~\cite{MestyanPozsgay2014} about $\beta=\mu=0$.
At $\Delta=1.5$, the resulting bound is
\begin{equation}
  \mathcal{B}\simeq0.0470,
  \label{eq:xxzNumericalBound}
\end{equation}
which is stable at the quoted precision under
increases in the numerical cutoffs.
Direct evaluation of Eq.~\eqref{eq:xxzGHDerror}
gives a maximum normalized profile discrepancy of
$0.0168$.
Both the mismatch and reservoir contrast are
proportional to $\delta\beta$, so the normalized
error need not vanish in linear response.

The finite-amplitude simulations use separately
prepared reservoirs with $\beta_{L,R}=\mp0.010$ and
$\mu_{L,R}=\mp0.10$.
The end-spin contributions to $\hat M_\alpha$ are
halved during preparation, and the junction bond is
switched on for evolution.
For the numerical joint-EOS readout, we use the mean
magnetization of the two sites forming each bond:
\begin{equation}
  e_{\rm snap,i}=e_{\rm eq}(\bar m_i,C_i),\qquad
  \bar m_i=\frac{m_i+m_{i+1}}2.
  \label{eq:xxzreadout}
\end{equation}

For Fig.~\ref{fig:xxz}(d), we average $m$ and $C$ over
$32768$ snapshots at each time before reconstructing
the energy. Spin transfer is summed over $64$ sites
on each side of the junction; energy transfer uses
all bonds except the junction. Both are normalized
by the initial difference between the reservoirs,
with energy evaluated using the same EOS.
We estimate sampling uncertainties by dividing the
snapshots into eight equal groups and repeating
the calculation with each group omitted in turn.
The initial values and EOS calibration are held
fixed, so the error bars account only for snapshot
sampling.

Finally, we note that the bound $\cal{B}$ varies smoothly with anisotropy, taking values in the range $\approx 1.4-6\%$ for $\Delta \in [1.1, 5]$. The choice of $\Delta=1.5$ therefore is not special.

\begin{figure}[b]
 \centering
 \includegraphics[width=\linewidth]{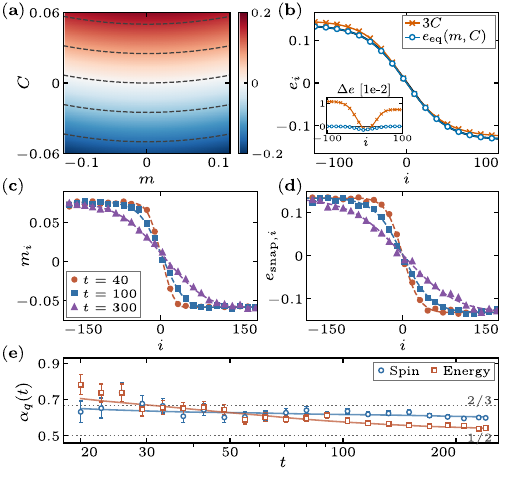}
 \caption{\textbf{Snapshot readout of spin and energy transport in the spin-1 Heisenberg chain.}
 \textbf{(a)}~Numerical EOS $e_{\rm eq}(m,C)$ from a periodic $L=8$ chain computed by exact diagonalization (colors); dashed contours show the second-order expression in Eq.~\eqref{eq:spin1twovariableeos}.
 \textbf{(b)}~Energy per site at $t=100$: directly computed energy (black), $3C$ from raw longitudinal correlations averaged over adjacent bonds (orange), and the full $(m,C)$ EOS readout (blue).
 Inset: each readout minus the directly computed site energy ($\Delta e$). 
  \textbf{(c)}-\textbf{(d)}~Magnetization and reconstructed energy profiles at $t=40,100,300$ (symbols, $N_{\rm shot}=32768$ per time), compared with directly computed profiles (dashed).
  Points are averaged over six sites, with $i=0$ denoting the initial wall center.
  \textbf{(e)}~Effective spreading exponents $\alpha_q(t)$ for spin and energy from snapshots (symbols) and direct simulation profiles (solid). The instantaneous exponents are estimated from fits over the time interval $[t/\sqrt2,\sqrt2\,t]$, with the gray dotted lines showing the reference exponents $1/2$ and $2/3$.
  Error bars show sampling uncertainties.
  The dynamics are simulated using DMT with bond dimension $\chi\le128$ on a chain of $L=512$ sites, with time step $\delta t=0.2$ and initial steps $\delta\beta=\delta\mu=0.1$. }
 \label{fig:spin1eos}
 \label{fig:spin1transport}
\end{figure}

\section{Spin and energy readout in the spin-1 Heisenberg chain}
\label{app:spin1}

We apply the joint spin and energy readout of Sec.~\ref{sec:xxz} to
the nonintegrable spin-$1$ Heisenberg chain,
\begin{equation}
 \hat H_\mathrm{XXX}^{S=1} =\sum_{i=-L/2+1}^{L/2-1}\hat h_i,\qquad
 \hat h_i=\hat{\mathbf S}_i\cdot\hat{\mathbf S}_{i+1}.
 \label{eq:spin1smhamiltonian}
\end{equation}
This model conserves energy and total magnetization.
Both are expected to exhibit diffusive transport at
asymptotically late times~\cite{Karadamoglou2004,Dupont2020},
although spin dynamics can remain close to superdiffusive
scaling over an extended time window~\cite{Dupont2020}.

We use the Gibbs-wall preparation of Sec.~\ref{sec:xxz}, now with
$\hat M_\alpha=\sum_{i\in\alpha}\hat S_i^z$ and
$\delta\beta=\delta\mu=0.1$.
For the uniform Gibbs averages, we write
$m=\langle\hat S_i^z\rangle$,
$C=\langle\hat S_i^z\hat S_{i+1}^z\rangle$, and
$e=\langle\hat h_i\rangle$.
At $\mu=0$, SU(2) invariance gives $e=3C$ at any
temperature.
For $\delta\mu=0$, this identity also holds on every
bond throughout the dynamics, since both the prepared
wall and the Hamiltonian are SU(2)-invariant.
The magnetization step therefore allows us to test
readout beyond this exact identity.

Near infinite temperature, $e_{\rm eq}(m,C)$ is analytic
and, by spin-flip symmetry, even in $m$.
At $\beta=0$, the Gibbs state factorizes, giving
$e=C=m^2$.
Together with the SU(2) relation, this fixes the leading
magnetization correction:
\begin{equation}
 e_{\rm eq}(m,C)=3C-2m^2+O(Cm^2,m^4).
 \label{eq:spin1twovariableeos}
\end{equation}
The nonzero slope
$\left.\partial_e C_{\rm eq}\right|_{e=m=0}=1/3$
ensures local invertibility of the energy readout.

We calibrate the full numerical EOS by exact
diagonalization of a periodic $L=8$ chain in sectors
of fixed total $S^z$.
Figure~\ref{fig:spin1eos}(a) compares this calibration
with Eq.~\eqref{eq:spin1twovariableeos}.
For the finite magnetization step,
Fig.~\ref{fig:spin1eos}(b) shows that $3C$ overestimates
the energy on the reservoir plateaus, while the full
$(m,C)$ EOS largely removes this offset.

For the snapshot reconstruction, we apply the numerical
EOS using the bond-averaged magnetization prescription
of Appendix~\ref{app:xxz}.
Adjacent bond-energy estimates are then averaged to
obtain the site-energy profile.
Figure~\ref{fig:spin1transport}(c)-(d) compare the
magnetization and reconstructed energy profiles with
those computed directly.
Both are recovered from the same $S^z$-basis snapshots.
To quantify spreading, we use the transfer lengths
$\ell_q(t)$ defined in Sec.~\ref{sec:xxz}, evaluated within mirrored
windows on either side of the junction, and estimate
$\alpha_q(t)=d\log\ell_q(t)/d\log t$ by local log--log fits.
The snapshot estimates follow the directly computed
exponents in Fig.~\ref{fig:spin1transport}(e).
Over the accessible time window, both exponents drift
downward: spin moves away from the $2/3$ reference
but remains above $1/2$, while energy approaches
$1/2$ more closely, consistent with previous findings~\cite{Dupont2020}.

\bibliography{ref}

@article{Bhakuni2024,
  author = {Bhakuni, D. S. and Verdel, R. and Muzzi, C. and Andreoni, R. and Aidelsburger, M. and Dalmonte, M.},
  title = {Diagnosing quantum transport from wave function snapshots},
  journal = {Phys. Rev. B},
  volume = {110},
  pages = {144204},
  year = {2024},
  doi = {10.1103/PhysRevB.110.144204},
  eprint = {},
  archivePrefix = {arXiv},
}

@misc{Yadav2026,
  author = {Yadav, D. and Bhakuni, D. S. and Agarwalla, B. K.},
  title = {Principal component analysis of wavefunction snapshots in non-equilibrium dynamics},
  year = {2026},
  eprint = {2604.01140},
  archivePrefix = {arXiv},
  primaryClass = {cond-mat.stat-mech},
}

@article{Elben2023,
  author = {Elben, A. and Flammia, S. T. and Huang, H.-Y. and Kueng, R. and Preskill, J. and Vermersch, B. and Zoller, P.},
  title = {The randomized measurement toolbox},
  journal = {Nat. Rev. Phys.},
  volume = {5},
  pages = {9},
  year = {2023},
  doi = {10.1038/s42254-022-00535-2},
  eprint = {},
  archivePrefix = {arXiv},
}

@article{Bernien2017,
  author = {Bernien, H. and Schwartz, S. and Keesling, A. and Levine, H. and Omran, A. and Pichler, H. and Choi, S. and Zibrov, A. S. and Endres, M. and Greiner, M. and Vuleti\'c, V. and Lukin, M. D.},
  title = {Probing many-body dynamics on a 51-atom quantum simulator},
  journal = {Nature},
  volume = {551},
  pages = {579},
  year = {2017},
  doi = {10.1038/nature24622},
  eprint = {},
  archivePrefix = {arXiv},
}

@article{Browaeys2020,
  author = {Browaeys, A. and Lahaye, T.},
  title = {Many-body physics with individually controlled {Rydberg} atoms},
  journal = {Nat. Phys.},
  volume = {16},
  pages = {132},
  year = {2020},
  doi = {10.1038/s41567-019-0733-z},
  eprint = {},
  archivePrefix = {arXiv},
}

@article{
Rosenberg2024,
author = {{E. Rosenberg \emph{et al.}}},
title = {Dynamics of magnetization at infinite temperature in a {Heisenberg} spin chain},
journal = {Science},
volume = {384},
number = {6691},
pages = {48-53},
year = {2024},
doi = {10.1126/science.adi7877},
URL = {https://www.science.org/doi/abs/10.1126/science.adi7877},
eprint = {}}

@article{Joshi2022,
  author = {Joshi, M. K. and Kranzl, F. and Schuckert, A. and Lovas, I. and Maier, C. and Blatt, R. and Knap, M. and Roos, C. F.},
  title = {Observing emergent hydrodynamics in a long-range quantum magnet},
  journal = {Science},
  volume = {376},
  pages = {720--724},
  year = {2022},
  doi = {10.1126/science.abk2400},
  eprint = {},
  archivePrefix = {arXiv},
}

@article{Wei2022,
  author = {Wei, D. and Rubio-Abadal, A. and Ye, B. and Machado, F. and Kemp, J. and Srakaew, K. and Hollerith, S. and Rui, J. and Gopalakrishnan, S. and Yao, N. Y. and Bloch, I. and Zeiher, J.},
  title = {Quantum gas microscopy of {Kardar-Parisi-Zhang} superdiffusion},
  journal = {Science},
  volume = {376},
  pages = {716},
  year = {2022},
  doi = {10.1126/science.abk2397},
  eprint = {},
  archivePrefix = {arXiv},
}

@article{Lesanovsky2012,
	Author = {Lesanovsky, Igor and Katsura, Hosho},
	Doi = {10.1103/PhysRevA.86.041601},
	Issue = {4},
	Journal = {Phys. Rev. A},
	Month = {Oct},
	Numpages = {5},
	Pages = {041601(R)},
	Publisher = {American Physical Society},
	Title = {Interacting {Fibonacci} anyons in a {Rydberg} gas},
	Url = {https://link.aps.org/doi/10.1103/PhysRevA.86.041601},
	Volume = {86},
	Year = {2012}}

@article{Ho2019,
  title = {Periodic Orbits, Entanglement, and Quantum Many-Body Scars in Constrained Models: Matrix Product State Approach},
  author = {Ho, Wen Wei and Choi, Soonwon and Pichler, Hannes and Lukin, Mikhail D.},
  journal = {Phys. Rev. Lett.},
  volume = {122},
  issue = {4},
  pages = {040603},
  numpages = {6},
  year = {2019},
  month = {Jan},
  publisher = {American Physical Society},
  doi = {10.1103/PhysRevLett.122.040603},
  url = {https://link.aps.org/doi/10.1103/PhysRevLett.122.040603}
}

@article{Khemani2019,
  title = {Signatures of integrability in the dynamics of {Rydberg}-blockaded chains},
  author = {Khemani, Vedika and Laumann, Chris R. and Chandran, Anushya},
  journal = {Phys. Rev. B},
  volume = {99},
  issue = {16},
  pages = {161101},
  numpages = {6},
  year = {2019},
  month = {Apr},
  publisher = {American Physical Society},
  doi = {10.1103/PhysRevB.99.161101},
  url = {https://link.aps.org/doi/10.1103/PhysRevB.99.161101}
}

@article{Bhakuni2026,
  title = {Anomalously fast transport in nonintegrable lattice gauge theories},
  author = {Bhakuni, Devendra Singh and Verdel, Roberto and Desaules, Jean-Yves and Serbyn, Maksym and Ljubotina, Marko and Dalmonte, Marcello},
  journal = {Phys. Rev. B},
  volume = {114},
  issue = {14},
  pages = {L140103},
  numpages = {8},
  year = {2026},
  month = {Sep},
  publisher = {American Physical Society},
  doi = {10.1103/3371-88bl},
  url = {https://link.aps.org/doi/10.1103/3371-88bl}
}

@article{Jiang2026,
  title = {Finite-temperature crossover from coherent magnons to energy superdiffusion in the PXP model},
  author = {Jiang, Shengtao and Desaules, Jean-Yves and Ljubotina, Marko and Scaffidi, Thomas},
  journal = {Phys. Rev. B},
  volume = {114},
  issue = {8},
  pages = {084306},
  numpages = {8},
  year = {2026},
  month = {Aug},
  publisher = {American Physical Society},
  doi = {10.1103/mx9k-1mc6},
  url = {https://link.aps.org/doi/10.1103/mx9k-1mc6}
}

@article{Kerschbaumer2025,
  title = {Quantum Many-Body Scars beyond the {PXP} Model in {Rydberg} Simulators},
  author = {Kerschbaumer, Aron and Ljubotina, Marko and Serbyn, Maksym and Desaules, Jean-Yves},
  journal = {Phys. Rev. Lett.},
  volume = {134},
  issue = {16},
  pages = {160401},
  numpages = {9},
  year = {2025},
  month = {Apr},
  publisher = {American Physical Society},
  doi = {10.1103/PhysRevLett.134.160401},
  url = {https://link.aps.org/doi/10.1103/PhysRevLett.134.160401}
}

@article{Omiya2022,
  title = {Quantum many-body scars in bipartite Rydberg arrays originating from hidden projector embedding},
  author = {Omiya, Keita and M\"uller, Markus},
  journal = {Phys. Rev. A},
  volume = {107},
  issue = {2},
  pages = {023318},
  numpages = {20},
  year = {2023},
  month = {Feb},
  publisher = {American Physical Society},
  doi = {10.1103/PhysRevA.107.023318},
  url = {https://link.aps.org/doi/10.1103/PhysRevA.107.023318}
}

@article{Choi2019,
	Author = {Choi, Soonwon and Turner, Christopher J. and Pichler, Hannes and Ho, Wen Wei and Michailidis, Alexios A. and Papi\ifmmode \acute{c}\else \'{c}\fi{}, Zlatko and Serbyn, Maksym and Lukin, Mikhail D. and Abanin, Dmitry A.},
	Doi = {10.1103/PhysRevLett.122.220603},
	Issue = {22},
	Journal = {Phys. Rev. Lett.},
	Month = {Jun},
	Numpages = {6},
	Pages = {220603},
	Publisher = {American Physical Society},
	Title = {Emergent SU(2) Dynamics and Perfect Quantum Many-Body Scars},
	Url = {https://link.aps.org/doi/10.1103/PhysRevLett.122.220603},
	Volume = {122},
	Year = {2019}}

@article{Lin2019exact,
  title = {Exact Quantum Many-Body Scar States in the Rydberg-Blockaded Atom Chain},
  author = {Lin, Cheng-Ju and Motrunich, Olexei I.},
  journal = {Phys. Rev. Lett.},
  volume = {122},
  issue = {17},
  pages = {173401},
  numpages = {5},
  year = {2019},
  month = {Apr},
  publisher = {American Physical Society},
  doi = {10.1103/PhysRevLett.122.173401},
  url = {https://link.aps.org/doi/10.1103/PhysRevLett.122.173401}
}

@article{Surace2020,
  title = {Lattice Gauge Theories and String Dynamics in {Rydberg} Atom Quantum Simulators},
  author = {Surace, Federica M. and Mazza, Paolo P. and Giudici, Giuliano and Lerose, Alessio and Gambassi, Andrea and Dalmonte, Marcello},
  journal = {Phys. Rev. X},
  volume = {10},
  issue = {2},
  pages = {021041},
  numpages = {14},
  year = {2020},
  month = {May},
  publisher = {American Physical Society},
  doi = {10.1103/PhysRevX.10.021041},
  url = {https://link.aps.org/doi/10.1103/PhysRevX.10.021041}
}

@article{Su2023,
title = {Observation of many-body scarring in a {Bose-Hubbard} quantum simulator},
  author = {Su, Guo-Xian and Sun, Hui and Hudomal, Ana and Desaules, Jean-Yves and Zhou, Zhao-Yu and Yang, Bing and Halimeh, Jad C. and Yuan, Zhen-Sheng and Papi\ifmmode \acute{c}\else \'{c}\fi{}, Zlatko and Pan, Jian-Wei},
  journal = {Phys. Rev. Res.},
  volume = {5},
  issue = {2},
  pages = {023010},
  numpages = {13},
  year = {2023},
  month = {Apr},
  publisher = {American Physical Society},
  doi = {10.1103/PhysRevResearch.5.023010},
  url = {https://link.aps.org/doi/10.1103/PhysRevResearch.5.023010}
}

@article{Bluvstein2021,
author = {D. Bluvstein  and A. Omran  and H. Levine  and A. Keesling  and G. Semeghini  and S. Ebadi  and T. T. Wang  and A. A. Michailidis  and N. Maskara  and W. W. Ho  and S. Choi  and M. Serbyn  and M. Greiner  and V. Vuletić  and M. D. Lukin },
title = {Controlling quantum many-body dynamics in driven {Rydberg} atom arrays},
journal = {Science},
volume = {371},
number = {6536},
pages = {1355-1359},
year = {2021},
doi = {10.1126/science.abg2530},
URL = {https://www.science.org/doi/abs/10.1126/science.abg2530}
}

@article{Zhao2024Observationquantumthermalizationrestricted,
  title = {Observation of Quantum Thermalization Restricted to {Hilbert} Space Fragments and $Z_{2k}$ Scars},
  author = {Zhao, Luheng and Datla, Prithvi Raj and Tian, Weikun and Aliyu, Mohammad Mujahid and Loh, Huanqian},
  journal = {Phys. Rev. X},
  volume = {15},
  issue = {1},
  pages = {011035},
  numpages = {18},
  year = {2025},
  month = {Feb},
  publisher = {American Physical Society},
  doi = {10.1103/PhysRevX.15.011035},
  url = {https://link.aps.org/doi/10.1103/PhysRevX.15.011035}
}

@article{FendleySachdev,
	Author = {Fendley, Paul and Sengupta, K. and Sachdev, Subir},
	Doi = {10.1103/PhysRevB.69.075106},
	Issue = {7},
	Journal = {Phys. Rev. B},
	Month = {Feb},
	Numpages = {15},
	Pages = {075106},
	Publisher = {American Physical Society},
	Title = {Competing density-wave orders in a one-dimensional hard-boson model},
	Url = {https://link.aps.org/doi/10.1103/PhysRevB.69.075106},
	Volume = {69},
	Year = {2004}}

@article{Ljubotina2023,
  author = {Ljubotina, M. and Desaules, J.-Y. and Serbyn, M. and Papi\'c, Z.},
  title = {Superdiffusive energy transport in kinetically constrained models},
  journal = {Phys. Rev. X},
  volume = {13},
  pages = {011033},
  year = {2023},
  doi = {10.1103/PhysRevX.13.011033},
  eprint = {},
  archivePrefix = {arXiv},
}

@misc{mishra2026theoryoutoftimeorderedtransport,
      title={Theory of Out-of-Time-Ordered Transport}, 
      author={Ruchira Mishra and Jiaozi Wang and Silvia Pappalardi and Luca V. Delacrétaz},
      year={2026},
      eprint={2512.11198},
      archivePrefix={arXiv},
      primaryClass={cond-mat.str-el},
      url={https://arxiv.org/abs/2512.11198}, 
}

@article{Schecter2018,
  title = {Many-body spectral reflection symmetry and protected infinite-temperature degeneracy},
  author = {Schecter, Michael and Iadecola, Thomas},
  journal = {Phys. Rev. B},
  volume = {98},
  issue = {3},
  pages = {035139},
  numpages = {8},
  year = {2018},
  month = {Jul},
  publisher = {American Physical Society},
  doi = {10.1103/PhysRevB.98.035139},
  url = {https://link.aps.org/doi/10.1103/PhysRevB.98.035139}
}

@article{Turner2018PRB,
  title = {Quantum scarred eigenstates in a {Rydberg} atom chain: Entanglement, breakdown of thermalization, and stability to perturbations},
  author = {Turner, C. J. and Michailidis, A. A. and Abanin, D. A. and Serbyn, M. and Papi{\ifmmode \acute{c}\else \'{c}\fi{}}, Z.},
  journal = {Phys. Rev. B},
  volume = {98},
  issue = {15},
  pages = {155134},
  numpages = {23},
  year = {2018},
  month = {Oct},
  publisher = {American Physical Society},
  doi = {10.1103/PhysRevB.98.155134},
  url = {https://link.aps.org/doi/10.1103/PhysRevB.98.155134}
}

@article{Wang2026ETH,
  title = {Eigenstate Thermalization Hypothesis Correlations via Nonlinear Hydrodynamics},
  author = {Wang, Jiaozi and Mishra, Ruchira and Yang, Tian-Hua and Delacr{\'e}taz, Luca V. and Pappalardi, Silvia},
  journal = {Phys. Rev. Lett.},
  volume = {136},
  issue = {13},
  pages = {130402},
  numpages = {10},
  year = {2026},
  month = {Mar},
  publisher = {American Physical Society},
  doi = {10.1103/prv4-948b},
  url = {https://link.aps.org/doi/10.1103/prv4-948b}
}

@article{Iadecola2019,
  title = {Quantum many-body scars from magnon condensation},
  author = {Iadecola, Thomas and Schecter, Michael and Xu, Shenglong},
  journal = {Phys. Rev. B},
  volume = {100},
  issue = {18},
  pages = {184312},
  numpages = {12},
  year = {2019},
  month = {Nov},
  publisher = {American Physical Society},
  doi = {10.1103/PhysRevB.100.184312},
  url = {https://link.aps.org/doi/10.1103/PhysRevB.100.184312}
}

@article{Kim2013,
  title = {Ballistic Spreading of Entanglement in a Diffusive Nonintegrable System},
  author = {Kim, Hyungwon and Huse, David A.},
  journal = {Phys. Rev. Lett.},
  volume = {111},
  issue = {12},
  pages = {127205},
  numpages = {5},
  year = {2013},
  month = {Sep},
  publisher = {American Physical Society},
  doi = {10.1103/PhysRevLett.111.127205},
  url = {https://link.aps.org/doi/10.1103/PhysRevLett.111.127205}
}

@article{Turner2018,
  author = {Turner, C. J. and Michailidis, A. A. and Abanin, D. A. and Serbyn, M. and Papi\'c, Z.},
  title = {Weak ergodicity breaking from quantum many-body scars},
  journal = {Nat. Phys.},
  volume = {14},
  pages = {745},
  year = {2018},
  doi = {10.1038/s41567-018-0137-5},
  eprint = {},
  archivePrefix = {arXiv},
}

@book{NielsenChuang2010, 
place={Cambridge}, 
title={Quantum Computation and Quantum Information: 10th Anniversary Edition}, 
publisher={Cambridge University Press}, 
author={Nielsen, Michael A. and Chuang, Isaac L.}, 
year={2010}
}

@article{Zotos1997,
  author = {Zotos, X. and Naef, F. and Prelov\v{s}ek, P.},
  title = {Transport and conservation laws},
  journal = {Phys. Rev. B},
  volume = {55},
  pages = {11029},
  year = {1997},
  doi = {10.1103/PhysRevB.55.11029},
  eprint = {},
  archivePrefix = {arXiv},
}

@article{KlumperSakai2002,
  author = {Kl\"umper, A. and Sakai, K.},
  title = {The thermal conductivity of the spin-1/2 {XXZ} chain at arbitrary temperature},
  journal = {J. Phys. A: Math. Gen.},
  volume = {35},
  pages = {2173},
  year = {2002},
  doi = {10.1088/0305-4470/35/9/307},
  eprint = {},
  archivePrefix = {arXiv},
}

@article{White2018,
  author = {White, C. D. and Zaletel, M. and Mong, R. S. K. and Refael, G.},
  title = {Quantum dynamics of thermalizing systems},
  journal = {Phys. Rev. B},
  volume = {97},
  pages = {035127},
  year = {2018},
  doi = {10.1103/PhysRevB.97.035127},
  eprint = {},
  archivePrefix = {arXiv},
}

@article{Rakovszky2022DAOE,
  author = {Rakovszky, T. and von Keyserlingk, C. W. and Pollmann, F.},
  title = {Dissipation-assisted operator evolution method for capturing hydrodynamic transport},
  journal = {Phys. Rev. B},
  volume = {105},
  pages = {075131},
  year = {2022},
  doi = {10.1103/PhysRevB.105.075131},
  eprint = {},
  archivePrefix = {arXiv},
}

@article{vonKeyserlingk2022Backflow,
  author = {von Keyserlingk, C. W. and Pollmann, F. and Rakovszky, T.},
  title = {Operator backflow and the classical simulation of quantum transport},
  journal = {Phys. Rev. B},
  volume = {105},
  pages = {245101},
  year = {2022},
  doi = {10.1103/PhysRevB.105.245101},
  eprint = {},
  archivePrefix = {arXiv},
}

@article{Srivatsa2026DAOE,
  author = {Srivatsa, N. S. and Lunt, O. and Rakovszky, T. and von Keyserlingk, C.},
  title = {Probing hydrodynamic crossovers with dissipation-assisted operator evolution},
  journal = {Phys. Rev. B},
  volume = {113},
  pages = {165150},
  year = {2026},
  doi = {10.1103/gz9n-v8ty},
  eprint = {},
  archivePrefix = {arXiv},
}

@article{YiThomas2024,
  author = {Yi-Thomas, S. and Ware, B. and Sau, J. D. and White, C. D.},
  title = {Comparing numerical methods for hydrodynamics in a one-dimensional lattice spin model},
  journal = {Phys. Rev. B},
  volume = {110},
  pages = {134308},
  year = {2024},
  doi = {10.1103/PhysRevB.110.134308},
  eprint = {},
  archivePrefix = {arXiv},
}

@article{CastroAlvaredo2016,
  author = {Castro-Alvaredo, O. A. and Doyon, B. and Yoshimura, T.},
  title = {Emergent hydrodynamics in integrable quantum systems out of equilibrium},
  journal = {Phys. Rev. X},
  volume = {6},
  pages = {041065},
  year = {2016},
  doi = {10.1103/PhysRevX.6.041065},
  eprint = {},
  archivePrefix = {arXiv},
}

@article{Bertini2016,
  author = {Bertini, B. and Collura, M. and De Nardis, J. and Fagotti, M.},
  title = {Transport in out-of-equilibrium {XXZ} chains: Exact profiles of charges and currents},
  journal = {Phys. Rev. Lett.},
  volume = {117},
  pages = {207201},
  year = {2016},
  doi = {10.1103/PhysRevLett.117.207201},
  eprint = {},
  archivePrefix = {arXiv},
}

@article{MestyanPozsgay2014,
  author = {Mesty{\'a}n, M. and Pozsgay, B.},
  title = {Short distance correlators in the {XXZ} spin chain for arbitrary string distributions},
  journal = {J. Stat. Mech.: Theory Exp.},
  volume = {2014},
  pages = {P09020},
  year = {2014},
  doi = {10.1088/1742-5468/2014/09/P09020},
  eprint = {},
  archivePrefix = {arXiv},
}

@article{Ljubotina2017,
  author = {Ljubotina, M. and \v{Z}nidari\v{c}, M. and Prosen, T.},
  title = {Spin diffusion from an inhomogeneous quench in an integrable system},
  journal = {Nat. Commun.},
  volume = {8},
  pages = {16117},
  year = {2017},
  doi = {10.1038/ncomms16117},
  eprint = {},
  archivePrefix = {arXiv},
}

@article{Ilievski2015,
  author = {Ilievski, E. and De Nardis, J. and Wouters, B. and Caux, J.-S. and Essler, F. H. L. and Prosen, T.},
  title = {Complete generalized {Gibbs} ensembles in an interacting theory},
  journal = {Phys. Rev. Lett.},
  volume = {115},
  pages = {157201},
  year = {2015},
  doi = {10.1103/PhysRevLett.115.157201},
  eprint = {},
  archivePrefix = {arXiv},
}

@article{Efron1979Bootstrap,
  author = {Efron, Bradley},
  title = {Bootstrap methods: another look at the jackknife},
  journal = {Ann. Statist.},
  volume = {7},
  number = {1},
  pages = {1--26},
  year = {1979},
  doi = {10.1214/aos/1176344552}
}

@Article{DePasquale2016,
author={De Pasquale, Antonella
and Rossini, Davide
and Fazio, Rosario
and Giovannetti, Vittorio},
title={Local quantum thermal susceptibility},
journal={Nature Communications},
year={2016},
month={Sep},
day={29},
volume={7},
number={1},
pages={12782},
issn={2041-1723},
doi={10.1038/ncomms12782},
url={https://doi.org/10.1038/ncomms12782}
}

@article{Capizzi2025,
  title = {Hydrodynamics and the Eigenstate Thermalization Hypothesis},
  author = {Capizzi, Luca and Wang, Jiaozi and Xu, Xiansong and Mazza, Leonardo and Poletti, Dario},
  journal = {Phys. Rev. X},
  volume = {15},
  issue = {1},
  pages = {011059},
  numpages = {21},
  year = {2025},
  month = {Mar},
  publisher = {American Physical Society},
  doi = {10.1103/PhysRevX.15.011059},
  url = {https://link.aps.org/doi/10.1103/PhysRevX.15.011059}
}

@article{Lux2014,
  title = {Hydrodynamic long-time tails after a quantum quench},
  author = {Lux, Jonathan and M{\"u}ller, Jan and Mitra, Aditi and Rosch, Achim},
  journal = {Phys. Rev. A},
  volume = {89},
  issue = {5},
  pages = {053608},
  numpages = {8},
  year = {2014},
  month = {May},
  publisher = {American Physical Society},
  doi = {10.1103/PhysRevA.89.053608},
  url = {https://link.aps.org/doi/10.1103/PhysRevA.89.053608}
}

@article{Pasqualetti2024,
  title = {Equation of State and Thermometry of the 2D $\mathrm{SU}(N)$ Fermi-Hubbard Model},
  author = {Pasqualetti, G. and Bettermann, O. and Darkwah Oppong, N. and Ibarra-Garc{\'{\i}}a-Padilla, E. and Dasgupta, S. and Scalettar, R. T. and Hazzard, K. R. A. and Bloch, I. and F{\"o}lling, S.},
  journal = {Phys. Rev. Lett.},
  volume = {132},
  issue = {8},
  pages = {083401},
  numpages = {8},
  year = {2024},
  month = {Feb},
  publisher = {American Physical Society},
  doi = {10.1103/PhysRevLett.132.083401},
  url = {https://link.aps.org/doi/10.1103/PhysRevLett.132.083401}
}

@article{Dixmerias2026,
  title = {Fluctuation Thermometry of an Atom-Resolved Quantum Gas: Beyond the Fluctuation-Dissipation Theorem},
  author = {Dixmerias, Maxime and Verstraten, Joris and Daix, Cyprien and Peaudecerf, Bruno and de Jongh, Tim and Yefsah, Tarik},
  journal = {Phys. Rev. Lett.},
  volume = {137},
  issue = {12},
  pages = {123401},
  numpages = {8},
  year = {2026},
  month = {Sep},
  publisher = {American Physical Society},
  doi = {10.1103/45fw-r1cp},
  url = {https://link.aps.org/doi/10.1103/45fw-r1cp}
}

@article{Hartke2020,
  title = {Doublon-Hole Correlations and Fluctuation Thermometry in a {Fermi}-{Hubbard} Gas},
  author = {Hartke, Thomas and Oreg, Botond and Jia, Ningyuan and Zwierlein, Martin},
  journal = {Phys. Rev. Lett.},
  volume = {125},
  issue = {11},
  pages = {113601},
  numpages = {6},
  year = {2020},
  month = {Sep},
  publisher = {American Physical Society},
  doi = {10.1103/PhysRevLett.125.113601},
  url = {https://link.aps.org/doi/10.1103/PhysRevLett.125.113601}
}

@article{McGinley2023,
  title = {Shadow Tomography from Emergent State Designs in Analog Quantum Simulators},
  author = {McGinley, Max and Fava, Michele},
  journal = {Phys. Rev. Lett.},
  volume = {131},
  issue = {16},
  pages = {160601},
  numpages = {7},
  year = {2023},
  month = {Oct},
  publisher = {American Physical Society},
  doi = {10.1103/PhysRevLett.131.160601},
  url = {https://link.aps.org/doi/10.1103/PhysRevLett.131.160601}
}

@article{Tran2023,
  title = {Measuring Arbitrary Physical Properties in Analog Quantum Simulation},
  author = {Tran, Minh C. and Mark, Daniel K. and Ho, Wen Wei and Choi, Soonwon},
  journal = {Phys. Rev. X},
  volume = {13},
  issue = {1},
  pages = {011049},
  numpages = {35},
  year = {2023},
  month = {Mar},
  publisher = {American Physical Society},
  doi = {10.1103/PhysRevX.13.011049},
  url = {https://link.aps.org/doi/10.1103/PhysRevX.13.011049}
}

@article{Hu2022,
  title = {Hamiltonian-driven shadow tomography of quantum states},
  author = {Hu, Hong-Ye and You, Yi-Zhuang},
  journal = {Phys. Rev. Res.},
  volume = {4},
  issue = {1},
  pages = {013054},
  numpages = {12},
  year = {2022},
  month = {Jan},
  publisher = {American Physical Society},
  doi = {10.1103/PhysRevResearch.4.013054},
  url = {https://link.aps.org/doi/10.1103/PhysRevResearch.4.013054}
}

@article{Bakr2009,
  author = {Bakr, W. S. and Gillen, J. I. and Peng, A. and F{\"o}lling, S. and Greiner, M.},
  title = {A quantum gas microscope for detecting single atoms in a {Hubbard}-regime optical lattice},
  journal = {Nature},
  volume = {462},
  pages = {74--77},
  year = {2009},
  doi = {10.1038/nature08482},
  eprint = {},
  archivePrefix = {arXiv}
}

@Article{Huang2020,
author={Huang, Hsin-Yuan
and Kueng, Richard
and Preskill, John},
title={Predicting many properties of a quantum system from very few measurements},
journal={Nature Physics},
year={2020},
month={Oct},
day={01},
volume={16},
number={10},
pages={1050-1057},
issn={1745-2481},
doi={10.1038/s41567-020-0932-7},
url={https://doi.org/10.1038/s41567-020-0932-7}
}

@article{Liu2024HamiltonianShadow,
  author = {Liu, Z. and Hao, Z. and Hu, H.-Y.},
  title = {Predicting arbitrary state properties from single {Hamiltonian} quench dynamics},
  journal = {Phys. Rev. Research},
  volume = {6},
  pages = {043118},
  year = {2024},
  doi = {10.1103/PhysRevResearch.6.043118},
  eprint = {},
  archivePrefix = {arXiv}
}

@article{Peruzzo2026,
  title = {Reconstructing quantum states and expectations via dynamical tomography},
  author = {Peruzzo, Marco and Grigoletto, Tommaso and Ticozzi, Francesco},
  journal = {Phys. Rev. A},
  volume = {113},
  issue = {3},
  pages = {032435},
  numpages = {18},
  year = {2026},
  month = {Mar},
  publisher = {American Physical Society},
  doi = {10.1103/x88t-j8k9},
  url = {https://link.aps.org/doi/10.1103/x88t-j8k9}
}

@article{ZanociSwingle2021,
  author = {Zanoci, C. and Swingle, B.},
  title = {Temperature-dependent energy diffusion in chaotic spin chains},
  journal = {Phys. Rev. B},
  volume = {103},
  pages = {115148},
  year = {2021},
  doi = {10.1103/PhysRevB.103.115148},
  eprint = {},
  archivePrefix = {arXiv},
}

@article{Rigol2008,
  author = {Rigol, M. and Dunjko, V. and Olshanii, M.},
  title = {Thermalization and its mechanism for generic isolated quantum systems},
  journal = {Nature},
  volume = {452},
  pages = {854--858},
  year = {2008},
  doi = {10.1038/nature06838},
  eprint = {},
  archivePrefix = {arXiv},
  primaryClass = {cond-mat.stat-mech},
}

@article{DAlessioRigol2014,
  author = {D'Alessio, L. and Rigol, M.},
  title = {Long-time behavior of isolated periodically driven interacting lattice systems},
  journal = {Phys. Rev. X},
  volume = {4},
  pages = {041048},
  year = {2014},
  doi = {10.1103/PhysRevX.4.041048},
  eprint = {},
  archivePrefix = {arXiv},
}

@article{Khemani2018,
  author = {Khemani, V. and Vishwanath, A. and Huse, D. A.},
  title = {Operator Spreading and the Emergence of Dissipative Hydrodynamics under Unitary Evolution with Conservation Laws},
  journal = {Phys. Rev. X},
  volume = {8},
  pages = {031057},
  year = {2018},
  doi = {10.1103/PhysRevX.8.031057},
  eprint = {},
  archivePrefix = {arXiv},
}

@article{Rakovszky2018,
  author = {Rakovszky, T. and Pollmann, F. and von Keyserlingk, C. W.},
  title = {Diffusive Hydrodynamics of Out-of-Time-Ordered Correlators with Charge Conservation},
  journal = {Phys. Rev. X},
  volume = {8},
  pages = {031058},
  year = {2018},
  doi = {10.1103/PhysRevX.8.031058},
  eprint = {},
  archivePrefix = {arXiv},
}

@article{Verstraete2004,
  author = {Verstraete, F. and Garc{\'i}a-Ripoll, J. J. and Cirac, J. I.},
  title = {Matrix Product Density Operators: Simulation of Finite-Temperature and Dissipative Systems},
  journal = {Phys. Rev. Lett.},
  volume = {93},
  pages = {207204},
  year = {2004},
  doi = {10.1103/PhysRevLett.93.207204},
  eprint = {},
  archivePrefix = {arXiv},
}

@article{ZwolakVidal2004,
  author = {Zwolak, M. and Vidal, G.},
  title = {Mixed-State Dynamics in One-Dimensional Quantum Lattice Systems: A Time-Dependent Superoperator Renormalization Algorithm},
  journal = {Phys. Rev. Lett.},
  volume = {93},
  pages = {207205},
  year = {2004},
  doi = {10.1103/PhysRevLett.93.207205},
  eprint = {},
  archivePrefix = {arXiv},
}

@article{Suzuki1990,
  author = {Suzuki, M.},
  title = {Fractal decomposition of exponential operators with applications to many-body theories and {Monte Carlo} simulations},
  journal = {Phys. Lett. A},
  volume = {146},
  number = {6},
  pages = {319--323},
  year = {1990},
  doi = {10.1016/0375-9601(90)90962-N},
}

@misc{Wang2026Hierarchical,
  author = {Wang, Jiaozi and Joshi, Manoj K. and Capizzi, Luca and Blatt, Rainer and Roos, Christian F. and Mazza, Leonardo and Poletti, Dario},
  title = {Observation of universal hierarchical relaxation in a quantum simulator},
  year = {2026},
  eprint = {2609.29817},
  archivePrefix = {arXiv},
  primaryClass = {cond-mat.stat-mech},
  doi = {10.48550/arXiv.2609.29817},
}

@article{Karadamoglou2004,
  author = {Karadamoglou, J. and Zotos, X.},
  title = {Diffusive Transport in Spin-1 Chains at High Temperatures},
  journal = {Phys. Rev. Lett.},
  volume = {93},
  pages = {177203},
  year = {2004},
  doi = {10.1103/PhysRevLett.93.177203},
  eprint = {},
  archivePrefix = {arXiv},
}

@article{Dupont2020,
  author = {Dupont, Maxime and Moore, Joel E.},
  title = {Universal spin dynamics in infinite-temperature one-dimensional quantum magnets},
  journal = {Phys. Rev. B},
  volume = {101},
  pages = {121106},
  year = {2020},
  doi = {10.1103/PhysRevB.101.121106},
  eprint = {},
  archivePrefix = {arXiv},
}

@article{Le2023Hydrodynamization,
  author = {Le, Yuan and Zhang, Yicheng and
            Gopalakrishnan, Sarang and
            Rigol, Marcos and Weiss, David S.},
  title = {Observation of hydrodynamization and local
           prethermalization in {1D} {Bose} gases},
  journal = {Nature},
  volume = {618},
  pages = {494--499},
  year = {2023},
  doi = {10.1038/s41586-023-05979-9}
}

@article{Kaufman2016Thermalization,
  author = {Kaufman, Adam M. and Tai, M. Eric and
            Lukin, Alexander and Rispoli, Matthew and
            Schittko, Robert and Preiss, Philipp M. and
            Greiner, Markus},
  title = {Quantum thermalization through entanglement
           in an isolated many-body system},
  journal = {Science},
  volume = {353},
  pages = {794--800},
  year = {2016},
  doi = {10.1126/science.aaf6725},
  eprint = {},
  archivePrefix = {arXiv}
}

@article{Nascimbene2010EOS,
  author = {Nascimb{\`e}ne, S. and Navon, N. and
            Jiang, K. J. and Chevy, F. and Salomon, C.},
  title = {Exploring the thermodynamics of a universal
           {Fermi} gas},
  journal = {Nature},
  volume = {463},
  pages = {1057--1060},
  year = {2010},
  doi = {10.1038/nature08814}
}

@article{Ku2012Thermodynamics,
  author = {Ku, Mark J. H. and Sommer, Ariel T. and
            Cheuk, Lawrence W. and Zwierlein, Martin W.},
  title = {Revealing the superfluid lambda transition
           in the universal thermodynamics of a
           unitary {Fermi} gas},
  journal = {Science},
  volume = {335},
  pages = {563--567},
  year = {2012},
  doi = {10.1126/science.1214987},
  eprint = {},
  archivePrefix = {arXiv}
}

@article{Kim2018DetailedBalance,
  author = {Kim, Hyosub and Park, YeJe and
            Kim, Kyungtae and Sim, H.-S. and
            Ahn, Jaewook},
  title = {Detailed Balance of Thermalization Dynamics
           in {Rydberg}-Atom Quantum Simulators},
  journal = {Phys. Rev. Lett.},
  volume = {120},
  pages = {180502},
  year = {2018},
  doi = {10.1103/PhysRevLett.120.180502}
}

@article{Shaw2025Ergodicity,
  author = {Shaw, Adam L. and Mark, Daniel K. and
            Choi, Joonhee and Finkelstein, Ran and
            Scholl, Pascal and Choi, Soonwon and
            Endres, Manuel},
  title = {Experimental Signatures of {Hilbert}-Space
           Ergodicity: Universal Bitstring Distributions
           and Applications in Noise Learning},
  journal = {Phys. Rev. X},
  volume = {15},
  pages = {031001},
  year = {2025},
  doi = {10.1103/h6xy-zpx4}
}

@article{Datla2026Statistical,
  author = {Datla, Prithvi Raj and Zhao, Luheng and
            Ho, Wen Wei and Klco, Natalie and Loh, Huanqian},
  title = {Statistical localization of {U(1)} lattice gauge theory
           in a Rydberg simulator},
  journal = {Nature Physics},
  volume = {22},
  pages = {355--361},
  year = {2026},
  doi = {10.1038/s41567-026-03183-w}
}

\end{document}